\documentclass[12pt, draftclsnofoot, onecolumn]{IEEEtran}
\usepackage[font=small,skip=0.5pt]{caption}
\usepackage{amsmath}
\usepackage{amssymb}
\usepackage{amsfonts}
\usepackage{amsbsy}
\usepackage{graphicx}

\usepackage{epstopdf}
\usepackage{mathtools}
\usepackage{etoolbox}
\usepackage[noadjust]{cite}
\makeatletter
\patchcmd{\@maketitle}
{\addvspace{0.5\baselineskip}\egroup}
{\addvspace{-1\baselineskip}\egroup}
{}
{}
\makeatother

\begin{document}
	
	\title{Overlay Satellite-Terrestrial Networks for IoT under Hybrid Interference Environments}
	
\author{\IEEEauthorblockN{Pankaj K. Sharma, Budharam Yogesh, Deepika Gupta, and Dong In Kim} 
	\thanks{Pankaj K. Sharma and Budharam Yogesh are with the Department of Electronics and Communication Engineering, National Institute of Technology Rourkela, Rourkela 769008, India. Email: \{sharmap, 218ec5123\}@nitrkl.ac.in.}
	\thanks{Deepika Gupta is with the Department of Electronics and Communication Engineering, Dr S P M International Institute of Information Technology, Naya Raipur, Naya Raipur 493661, India. Email: deepika@iiitnr.edu.in.}
	\thanks{Dong In Kim is with the Department of Electrical and Computer Engineering, Sungkyunkwan University, Suwon, South Korea. Email: dikim@skku.ac.kr.}
}
	\maketitle
	
	\begin{abstract}
		In this paper, we consider an overlay satellite-terrestrial network (OSTN) where an opportunistically selected terrestrial internet-of-things (IoT) network assists the primary satellite communications as well as accesses the spectrum for its own communications under hybrid interference received from extra-terrestrial sources (ETSs) and terrestrial sources (TSs). Herein, the IoT network adopts power-domain multiplexing to amplify-and-forward the superposed satellite and IoT signals. Considering a unified analytical framework for shadowed-Rician fading with integer/non-integer Nakagami-\emph{m} parameter for satellite and interfering ETSs links along with the integer/non-integer Nakagami-\emph{m} fading for terrestrial IoT and interfering TSs links, we derive the outage probability (OP) of both satellite and IoT networks. Further, we derive the respective asymptotic OP expressions to reveal the diversity order of both satellite and IoT networks under the two conditions, namely when the transmit power of interferers: $(a)$ remains fixed; and $(b)$ varies proportional to the transmit powers of main satellite and IoT users. We show that the proposed OSTN with adaptive power-splitting factor benefits the IoT network while guaranteeing certain quality-of-service (QoS) of satellite network. We verify the numerical results by simulations.
	\end{abstract}
	
	
	\section{Introduction}
	\IEEEPARstart{I}{ntegration} of cooperative relaying to satellite networks has recently been emerged as a popular paradigm for reliable communications between a satellite and terrestrial user equipment (UE), especially when the satellite-terrestrial UE link is severely masked \cite{evans}, \cite{chini2010survey} (i.e., in the presence of heavy clouds, physical blockages, ground user in tunnels, etc.). Consequently, the dual hop satellite-terrestrial networks (STNs) with integrated terrestrial relay infrastructure have been evolved and received tremendous research interests. The STNs are mainly implemented in an integrated and hybrid manners \cite{jo2011satellite}, \cite{sreng}. While the integrated STNs utilize the same spectrum resources for communication over both the satellite-to-relay and relay-to-UE hops, the hybrid STNs make use of different spectrum resources for communication over these two hops. Hence, in integrated STNs, the terrestrial nodes may subject to hybrid interference from co-channel extra-terrestrial sources (ETSs) as well as terrestrial sources (TSs). In general, the satellite and terrestrial links in STNs are subject to shadowed-Rician (SR) and Nakagami-\emph{m} fading, respectively. 
	
	Most recently, a terrestrial ecosystem of extraordinarily large number of wirelessly connected devices pertaining to numerous applications, e.g., home appliances, vehicles, industrial sensors, etc., known as internet-of-things (IoT), has been evolved \cite{iot}. Intuitively, these billions of IoT devices are expected to increase tremendously the demand for spectrum resources in upcoming years. To this end, cognitive radio may be envisioned as a viable solution to deal with such spectrum scarcity in future. Cognitive radio enables the sharing of licensed spectrum of a primary network with an unlicensed secondary network as long as the quality-of-service (QoS) requirements of the primary network are protected \cite{haykin2005cognitive}. Most popular cognitive radio models are the underlay and overlay \cite{manna}, \cite{zou2010adaptive}. In the underlay model, the transmit power of secondary devices is strictly constrained to safeguard the primary network from harmful interference. On the contrary, in the overlay model, the secondary devices cooperatively assist the primary communications alongside their own secondary communications based on a less restrictive power-domain multiplexing of primary and secondary signals. Cognitive radio has recently been incorporated to STNs for higher spectral efficiency \cite{jia2016broadband}-\nocite{*}\cite{sksharma}.     
	 In view of evergrowing IoT applications, an overlay satellite-terrestrial network (OSTN) is of great interest where the primary satellite spectrum (e.g., direct-to-home television bands, etc.) can be shared with secondary IoT devices. Herein, an IoT device by taking the role of a cooperative node not only can access the primary satellite spectrum for its own communications, but can also enhance the reliability of satellite communications based on cooperative diversity. It is worth mentioning that the integration of IoT to low earth orbit (LEO)/high earth orbit (HEO) satellite systems has already been envisioned in existing standards, e.g., digital video broadcast-satellite-second generation extension (DVB-S$2$X) \cite{dvb}. Vodafone along with Inmarsat have started to integrate satellites with terrestrial IoT devices to increase the cellular connectivity \cite{voda}. SpaceX organization has started Starlink project for satellite-terrestrial internet facilities \cite{star}. However, with the densification of STNs aiming futuristic sixth-generation ($6$G) services, the hybrid interference originating from both ETSs and TSs would be inevitable. Note that the common interfering ETSs include multiple co-channel beams-based satellites, co-channel satellites, and modern high altitude platforms (HAPs) (i.e., balloons, unmanned aerial vehicles (UAVs), etc.). Whereas, the common TSs include cellular systems, device-to-device (D2D), etc. 
   
	\subsection{Prior Arts, Motivation, and Contributions}
	The performance of STNs has been actively investigated in literature by taking into account decode-and-forward (DF) \cite{sreng}, \cite{mratdf} and amplify-and-forward (AF) \cite{bhatna}-\cite{pku} relays. The work in \cite{huaq} has investigated the performance of an integrated STN. In \cite{miridakis2014dual}, the performance of dual-hop multi-antenna STNs has been analyzed. Most recently, STNs with mobile unmanned aerial vehicle (UAV) relaying has been investigated in \cite{pksu}. The works in \cite{kanmlin} and \cite{pansec} have investigated the secrecy performance of STNs with ground and UAV relays, respectively. Further, the performance of cognitive STNs has been analyzed in \cite{guo}-\cite{lagunas2015resource}. Specifically, the performance of underlay STNs has been investigated in \cite{guo}, \cite{kola} whereas the performance of OSTNs has been analyzed in \cite{pkss}. The optimal resource allocation for cognitive STNs has been considered in \cite{vassaki2013power}, \cite{lagunas2015resource}. The authors in \cite{kangdf} and \cite{kangc} have investigated the performance of single- and multi-antenna STNs with interference from TSs, respectively. The performance of AF-based STNs for generalized integer and non-integer SR fading has been analyzed in \cite{yang2015performance} under terrestrial interference. Moreover, the outage performance of multiuser STNs with imperfect channel state information (CSI) was assessed in \cite{pkueucnc} in the presence of interfering TSs. The outage performance of integrated STNs with terrestrial interferers has been analyzed in \cite{ruanli}. Furthermore, the work in \cite{guo2018performance} has investigated the performance of STNs with hardware impairments and interference. On another hand, the outage performance of underlay cognitive STNs under interfering TSs has been investigated in \cite{kangtvtc}. Note that the majority of aforementioned works have considered the performance analysis of STNs with interference from TSs only by neglecting the crucial interference from ETSs. More importantly, very few works have considered the analytical framework for SR fading with integer (INT)/non-integer (NINT) Nakagami-\emph{m} parameter for satellite and interfering ETSs links along with the INT/NINT Nakagami-\emph{m} fading for terrestrial IoT and interfering TSs links.
	It is quite intuitive that the interference originating from ETSs may have significant impact on the performance of STNs since SR fading is dominated by the line-of-sight (LoS) propagation. So far, in the context of cognitive STNs, the performance analysis of OSTNs taking into account the hybrid interference from ETSs and TSs has not been addressed in open literature. An initial attempt in this direction is made for interference-limited IoT-enabled OSTN in \cite{pksnc} by considering only integer SR and Nakagami-\emph{m} fading parameters. Furthermore, a unified framework for INT/NINT Nakagami-\emph{m} parameter-based SR fading for main satellite and interfering ETSs links has not been introduced in previous works.   
	
	Motivated by the above, in this paper, we investigate the outage performance of an IoT-enabled OSTN where an opportunistically-selected IoT network assists the primary satellite communications while communicating with its own receiver in the presence of hybrid interference from ETSs and TSs. We consider a unified analytical framework that enables the evaluation of system performance under SR fading with INT/NINT Nakagami-\emph{m} parameter for main satellite and interfering ETSs links. We further consider the INT/NINT Nakagami-\emph{m} fading for IoT and interfering TSs links. 
	The main contributions of the paper can be summarized as follows:
	\begin{itemize}
		\item We consider a unified probability density function (pdf)-based analytical framework for SR fading with INT/NINT Nakagami-\emph{m} parameter for satellite and interfering ETSs links. We further consider the INT/NINT Nakagami-\emph{m} fading for IoT and interfering TSs links. Based on the above, we present the statistical characterization of the independent and non-identically distributed (i.ni.d) hybrid interference from ETSs and TSs. 
		\item Based on the proposed statistical characterization of hybrid interference, we derive the OP of the satellite and IoT networks of the considered OSTN for various combinations of INT/NINT SR and Nakagami-\emph{m} fading scenarios. In particular, for non-integer cases of SR and terrestrial Nakagami-\emph{m} fading, the proposed analytical solution is presented in terms of convergent infinite series whose tightness under certain finite truncation of terms is depicted in numerical results.
		\item We derive the corresponding asymptotic OP expressions for INT/NINT scenarios of both SR and terrestrial Nakagami-\emph{m} fading. We assess the achievable diversity order of the satellite and IoT networks under the following conditions, namely when the transmit power of interferers $(a)$ remains fixed; and $(b)$ varies proportionally with the transmit power of main satellite and IoT users. 
		\item We consider the fixed as well as adaptive power-splitting factors to compare the performance of satellite and IoT networks of the OSTN. Nevertheless, we depict the impact of various system and channel parameters on the performance of considered OSTN.  
	\end{itemize}
	
	The rest of the paper is organized as follows: In Section \ref{sysmod},
	we detail the system, channel, and propagation models. We also describe the considered opportunistic IoT network selection strategy. Section \ref{statc} presents the statistical characterization of combined extra-terrestrial and terrestrial interference. Sections \ref{per} and \ref{sec} present the outage performance analysis of satellite and IoT networks, respectively. Section \ref{psf} formulates the adaptive power-splitting factor. Section \ref{num} presents the numerical and simulation results, and finally, the conclusions are drawn in Section \ref{con}.
	
	\textit{Notations:} Throughout the paper $f_X(x)$ denotes the pdf of random variable $X$. The cdf stands for the probability density function cumulative distribution function $F_X(x)$ of $X$. The functions $\Gamma(\cdot)$, $\Upsilon(\cdot,\cdot)$, and $\Gamma(\cdot,\cdot)$ are the gamma, lower incomplete gamma, and upper incomplete gamma functions, respectively. $\mathbb{E}[\cdot]$ denotes the statistical expectation.
	\section{System Description}\label{sysmod}
	\subsection{System Model}
\begin{figure}[!t]
		\centering
		\includegraphics[width=2.5in]{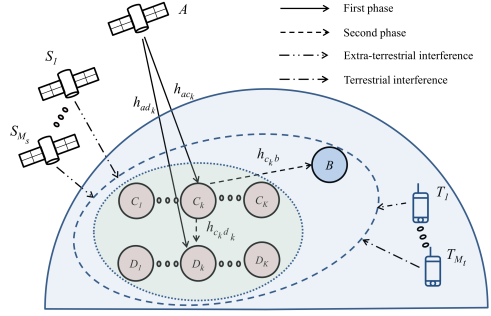}
		\vspace{5pt}
		\caption{OSTN with interfering ETSs and TSs.}
		\label{system}
	\end{figure}

	As shown in Fig. \ref{system}, we consider an OSTN comprising of a primary satellite transmitter ($A$)-terrestrial receiver ($B$) pair and multiple secondary IoT transmitter ($C_{k}$)-receiver ($D_{k}$) pairs, $k=1,...,K$. In addition, we consider that the cluster of multiple secondary IoT transmitter-receiver pairs comprising of IoT transmitters $\{C_k\}_{k=1}^{K}$ and the IoT receivers $\{D_k\}_{k=1}^{K}$ along with the receiver $B$ are inflicted by $M_s$ extra-terrestrial satellite interferers $\{S_j\}_{j=1}^{M_s}$ and $M_t$ terrestrial interferers $\{T_l\}_{l=1}^{M_t}$,  
	respectively. We assume that the direct link between satellite $A$ and its receiver $B$ is masked due to severe shadowing, blocking, etc. Herein, the secondary IoT transmitters compete to utilize the primary satellite network's spectrum in lieu of opportunistically assisting the satellite-to-ground communications based on the overlay spectrum sharing principle. According to the overlay principle, a selected secondary IoT transmitter $C_{k}$ serves as a relay that splits its total transmit power $P_c$ to multiplex the received primary signal and its own secondary signal in power domain with power levels $\mu P_c$ and $(1-\mu)P_c$, respectively, where $\mu\in(0,1)$. The channels pertaining to the links $A\rightarrow C_k$, $C_k\rightarrow B$, and $C_k\rightarrow D_k$ are denoted as $h_{ac_k}$, $h_{c_kb}$, and $h_{c_kd_k}$, respectively.  Also, $\{h_{sj}\}_{j=1}^{M_{s}}$ and $\{h_{tl}\}_{l=1}^{M_{t}}$ represent the channels from $S_j$ and $T_l$ to the cluster of all IoT transmitter-receiver pairs $C_k-D_k$, $k=1,...,K$. The thermal noise at each receiver node is assumed to be additive white Gaussian noise (AWGN) with mean zero and variance $\sigma^2$.
	
	\subsection{Propagation Model}
	The overall communication from satellite $A$ to terrestrial receiver $B$ takes place in two consecutive time phases with the help of a selected AF IoT transmitter relay $C_k$. While assisting the primary satellite communications, the IoT transmitter $C_k$ simultaneously communicates with IoT receiver $D_k$. 
	
	In the first phase, the satellite $A$ transmits a unit energy signal $x_{a}$ to IoT transmitter $C_{k}$ with transmit power $P_{a}$, which is also received by the IoT receiver $D_{k}$. Thus, the signals received at $C_{k}$ and $D_{k}$ can be expressed as
	\begin{align}\label{asrd}
	y_{ai}&=\sqrt{P_{a}}h_{ai}x_{a}+{I}_{s}+{I}_{t}+n_{ai},
	\end{align}
	where $i\in \{c_k,d_k\}$, $I_{s}=\sum_{j=1}^{M_s}\sqrt{P_{s}}h_{sj}x_j$ and $I_{t}=\sum_{l=1}^{M_t}\sqrt{P_{t}}h_{tl}x_l$ are the interferences received from extra-terrestrial and terrestrial interferers with respective transmit powers $P_s$ and $P_t$, respectively, and $n_{ai}$ is the AWGN. 
	
	In the second phase, the IoT transmitter $C_{k}$ combines the amplified primary signal $y_{ac_{k}}$ and its own secondary signal $x_{c_{k}}$ using superposition coding by splitting its total power as $\mu P_c$ and $(1-\mu)P_c$ among these signals, respectively. The resulting network-coded signal can be given as 
	\begin{align}\label{comb}
	z_{c_{k}}=\sqrt{\mu P_{c}}\frac{y_{ac_{k}}}{\sqrt{|y_{ac_{k}}|^{2}}}+\sqrt{(1-\mu) P_{c}}x_{c_{k}},
	\end{align}
	where $\mu\in (0,1)$ is a power-splitting factor. The IoT transmitter $C_k$ then broadcasts the superposed signal $z_{c_{k}}$ which is received by the nodes $B$ and $D_k$. The received signals at $B$ and $D_k$ are given, respectively, as     
	\begin{align}
	y_{c_{k}\upsilon}&=h_{c_{k}\upsilon}z_{c_{k}}+{I}_{s}+{I}_{t}+n_{c_{k}\upsilon},\label{asro}
	\end{align}
	where $\upsilon\in \{b,d_k\}$, $I_{s}$ and $I_t$ remain the same as defined previously, and $n_{c_{k}\upsilon}$ is the AWGN. Thus, the signal-to-interference-plus-noise ratio (SINR) at $B$ via relay link is given by
	\begin{align}\label{snrs}
	\Lambda_{ac_{k}b}&=\frac{\mu\hat{\Lambda}_{ac_{k}}\hat{\Lambda}_{c_{k}b}}{(1-\mu)\hat{\Lambda}_{ac_{k}}\hat{\Lambda}_{c_{k}b}+\hat{\Lambda}_{ac_{k}}+\hat{\Lambda}_{c_{k}b}+1},
	\end{align}
	where $\hat{\Lambda}_{ac_{k}}=\frac{{\Lambda}_{ac_{k}}}{W_c+1}$, $\hat{\Lambda}_{c_{k}b}=\frac{{\Lambda}_{c_{k}b}}{W_c+1}$, $\Lambda_{ac_{k}}=\eta_{a}|h_{ac_{k}}|^{2}$, $\eta_{a}=\frac{P_{a}}{\sigma^{2}}$, $W_c\triangleq W_s+W_t$, $W_s=\sum_{j=1}^{M_s}\Lambda_{sj}$, $\Lambda_{sj}=\eta_{s}|h_{sj}|^2$, $\eta_s=\frac{{P_{s}}}{\sigma^2}$, $W_t= \sum_{l=1}^{M_t}\Lambda_{tl}$, $\Lambda_{tl}=\eta_t|h_{tl}|^2$, $\eta_t=\frac{{P_{t}}}{\sigma^2}$, $\Lambda_{c_{k}b}=\eta_{c}|h_{c_{k}b}|^{2}$, and $\eta_{c}=\frac{P_{c}}{\sigma^{2}}$. 
	
	Moreover, from (\ref{asro}), we observe that the received signal $y_{c_{k}d_{k}}$ at $D_{k}$ contains the primary satellite signal $x_a$ which can be cancelled\footnote{Although the assumption of perfect successive interference cancellation at $D$ may be idealistic, it is followed for analytical tractability. It may be true for high signal-to-noise ratio (SNR) transmissions and/or very good channel conditions, e.g., clear sky, etc. However, the situations may arise when primary link is blocked with non-zero probability at $D_k$. The analysis for this case is mathematically tedious and may be deferred to future works.} by $D_k$ since a copy of $x_a$ is already received by it in the first phase. Hence, the equivalent SINR at the IoT receiver $D_{k}$ after primary interference cancellation is given by
	\begin{align}\label{secb}
	\Lambda_{ac_{k}d_{k}}&=\frac{(1-\mu)\hat{\Lambda}_{c_{k}d_{k}}(\hat{\Lambda}_{ac_{k}}+1)}{\mu\hat{\Lambda}_{c_{k}d_{k}}+\hat{\Lambda}_{ac_{k}}+1},
	\end{align}
	where $\hat{\Lambda}_{c_{k}d_{k}}=\frac{{\Lambda}_{c_{k}d_{k}}}{W_c+1}$ and ${\Lambda}_{c_{k}d_{k}}=\eta_{c}|h_{c_{k}d_{k}}|^{2}$.
	
	\subsection{Criteria for IoT Network Selection}
	We now discuss the criteria for selection of the best IoT network (i.e., selection of the pair $C_{k^{\ast}}$$-$$D_{k^{\ast}}$).
	To provide opportunistic spectrum access to IoT network, the best secondary network should be the one that incentivizes the satellite network with reliable communication. Hereby, if perfect CSI is available for  $A$$-$$C_{k}$ and $C_{k}$$-$$B$ link, an opportunistic strategy can be employed to maximize the end-to-end SINR at the satellite receiver $B$ as
	\begin{align}\label{ous}
	k^{\ast}=\arg \displaystyle \max_{k\in\{1,...,K\}} \Lambda_{ac_{k}b}.
	\end{align}
	Note that the aforementioned strategy can be implemented in both the centralized and distributed manner \cite{bletsas2006simple}. Hereby, we assume that perfect CSI is available for required IoT network selection. However, it is intricate to obtain perfect CSI in STNs due to large distance and delay between satellite and ground UE. Although some works \cite{irvine1992symbol}-\cite{mkar2} have recently dealt with the issue of CSI acquisition in STNs, it remains an open research problem of great interest.

	\subsection{Channel Models}\label{srf}
	\subsubsection{Main Satellite and Extra-Terrestrial Interference Channels}
	The channels $h_{ai_k}$, $i\in \{c,d\}$, $k\in\{1,...,K\}$ pertaining to main satellite links follow the shadowed-Rician (SR) fading. Consequently, the probability density function (pdf) of independently and identically distributed (i.i.d.) squared channels $|h_{ai_k}|^{2}$ can be given by
	\begin{align}\label{pdfrsaa}
	f_{|h_{ai_k}|^{2}}(x)&=\alpha_{ai} \textmd{e}^{-\beta_{ai}x}\,_1F_1(m_{ai};1;\delta_{ai}x), x\geq0,
	\end{align}
	where $\alpha_{ai}=(2\flat_{ai} m_{ai}/(2\flat_{ai} m_{ai}+\Omega_{ai}))^{m_{ai}}/2\flat_{ai} $, $\beta_{ai}=1/2\flat_{ai} $, $\delta_{ai}=\Omega_{ai}/(2\flat_{ai} )(2\flat_{ai} m_{ai}+\Omega_{ai})$, $\Omega_{ai}$ and $2\flat_{ai} $ are, respectively, the average power of the LoS and multipath components, $m_{ai}$ denotes the fading severity, and $_1F_1(\cdot;\cdot;\cdot)$ is the confluent hypergeometric function of the first kind \cite[eq. 9.210.1]{grad}. In general, the pdf in (\ref{pdfrsaa}) can be re-expressed under the two cases based on the values of parameter $m_{ai}$ being INT or NINT. As such, the corresponding pdfs can be represented in a unified form as \cite{miridakis2014dual}, \cite{guo2018performance}, \cite{abdi2003new} 
	\begin{align}\label{pdfaa}
	f_{|h_{ai_k}|^{2}}(x)&=\alpha_{ai} \sum_{\kappa=0}^{\varpi_{(\nu,ai)}}\zeta_{(\nu,ai)}(\kappa)x^{\kappa}\textmd{e}^{-\beta_{(\nu,ai)}x},
	\end{align}
	where $\nu\in\{\textmd{INT},\textmd{NINT}\}$. It follows that $\varpi_{(\textmd{INT},ai)}=m_{ai}-1$, $\zeta_{(\textmd{INT},ai)}(\kappa)=(-1)^{\kappa}(1-m_{ai})_{\kappa}\delta_{ai}^{\kappa}/(\kappa!)^{2}$, $\beta_{(\textmd{INT},ai)}=\beta_{ai}-\delta_{ai}$, for $\nu=\textmd{INT}$, and $\varpi_{(\textmd{NINT},ai)}=\infty$, $\zeta_{(\textmd{NINT},ai)}(\kappa)=(m_{ai})_{\kappa}\delta_{ai}^{\kappa}/(\kappa!)^{2}$, $\beta_{(\textmd{NINT},ai)}=\beta_{ai}$, for $\nu=\textmd{NINT}$, with $(\cdot)_{\kappa}$ as the Pochhammer symbol \cite[p. xliii]{grad}. Here, by making a transformation of variable, the pdf of the random variable $\Lambda_{ai_k}=\eta_{a}|h_{ai_k}|^{2}$ can be expressed as
	\begin{align}\label{pdfrs}
	f_{\Lambda_{ai_k}}(x)=\alpha_{ai} \sum_{\kappa=0}^{\varpi_{(\nu,ai)}}\frac{\zeta_{(\nu,ai)}(\kappa)}{(\eta_{a})^{\kappa+1}}x^{\kappa}
	\textmd{e}^{-\frac{\beta_{(\nu,ai)}}{\eta_{a}}x}.
	\end{align}
	The corresponding cumulative distribution function (cdf) $F_{\Lambda_{ai_k}}(x)$ can be computed, by integrating the result in (\ref{pdfrs}) with the aid of \cite[eq. 3.351.2]{grad}, as
	\begin{align}\label{pz}
	F_{\Lambda_{ai_k}}(x)&=1-\alpha_{ai}\sum_{\kappa=0}^{\varpi_{(\nu,ai)}}\frac{\zeta_{(\nu,ai)}(\kappa)}{(\eta_{a})^{\kappa+1}}\sum_{p=0}^{\kappa} \frac{\kappa!}{p!} \left(\frac{\beta_{(\nu,ai)}}{\eta_{a}}\right)^{-(\kappa+1-p)}x^{p}\textmd{e}^{-\frac{\beta_{(\nu,ai)}}{\eta_{a}}x}.
	\end{align}
	
	Likewise, we can obtain the pdf of i.i.d. squared interferers' channels $|h_{sj}|^{2}$ (i.e., $f_{|h_{sj}|^{2}}(x)$) using (\ref{pdfaa}) for corresponding INT and NINT cases of SR fading by replacing the parameters $\{\varpi_{(\nu,ai)}, \zeta_{(\nu,ai)}(\kappa), \beta_{(\nu,ai)}\}$ and $\{\alpha_{ai}, \beta_{ai}, \flat_{ai}, \delta_{ai}, m_{ai}, \Omega_{ai}\}$ by $\{\varpi_{(\nu,s)}, \zeta_{(\nu,s)}(\kappa), \beta_{(\nu,s)}\}$ and\\ $\{\alpha_{s}, \beta_{s}, \flat_{s}, \delta_{s}, m_{s}, \Omega_{s}\}$, respectively, for $j=1,...,M_s$. Consequently, the pdf $f_{\Lambda_{sj}}(x)$ and cdf $F_{\Lambda_{sj}}(x)$ for random variable $\Lambda_{sj}=\eta_{s}|h_{sj}|^{2}$ can be obtained with the help of (\ref{pdfrs}) and (\ref{pz}), respectively, based on the aforementioned procedure and $\eta_a$ replaced by $\eta_s$.    
	\subsubsection{Main Terrestrial and Terrestrial Interference Channels}
	
	The channels for main terrestrial links (i.e., $h_{c_{k}\upsilon}$, $\upsilon\in \{b,d_{k}\}$) and terrestrial interferers follow Nakagami-\emph{m} fading. Accordingly, the pdf $f_{\Lambda_{c_k\upsilon}}(x)$ and cdf $F_{\Lambda_{c_k\upsilon}}(x)$ of i.i.d. terrestrial links $\Lambda_{c_k\upsilon}=\eta_c|h_{c_k\upsilon}|^2$ can be obtained by the transformation of gamma variates, respectively, as
	\begin{align}\label{nak}
	f_{\Lambda_{c_k\upsilon}}(x)&=\left( \frac{m_{c\upsilon}}{\Omega_{c\upsilon}\eta_c}\right)^{m_{c\upsilon}} \frac{x^{m_{c\upsilon}-1}}{\Gamma(m_{c\upsilon})}\,\textmd{e}^{ -\frac{m_{c\upsilon}}{\Omega_{c\upsilon}\eta_c} x}
	\end{align}
	and
	\begin{align}\label{nakq}
	F_{\Lambda_{c_k\upsilon}}(x)&=\frac{1}{\Gamma(m_{c\upsilon})}\Upsilon \left(m_{c\upsilon}, \frac{m_{c\upsilon} x}{\Omega_{c\upsilon}\eta_{c}}\right),
	\end{align}
	where $\upsilon\in \{b,d\}$, $k=1,...,K$, $m_{cv}$ and $\Omega_{c\upsilon}$ is the fading severity and average fading power, respectively, with $\Upsilon(\cdot,\cdot)$ and $\Gamma(\cdot)$ as the lower incomplete and the complete gamma functions \cite[eqs. 8.310.1 and 8.350.1]{grad}, respectively. 
	
	Further, the pdf $f_{\Lambda_{c_k\upsilon}}(x)$ and cdf $F_{\Lambda_{c_k\upsilon}}(x)$ of i.i.d. random variables $\Lambda_{tl}=\eta_{t}|h_{tl}|^{2}$ corresponding to terrestrial interference links can be, respectively, expressed using (\ref{nak}) and (\ref{nakq}) by replacing therein the parameters $\{m_{c\upsilon},\Omega_{c\upsilon},\eta_c\}$ with  $\{m_{t},\Omega_{t},\eta_t\}$, for $l=1,...,M_t$.  
	
	Note that based on the values taken by the parameters $\{m_{ai},m_s\}$ pertaining to the main satellite and extra-terrestrial interference channels and $\{m_{c\upsilon},m_t\}$ corresponding to the main terrestrial and terrestrial interference channels, various scenarios can be analyzed as shown in Table \ref{tab}.  
	\begin{table}[t]
		\renewcommand{\arraystretch}{1.3}
		\caption{Illustration of various scenarios based upon the values taken by $\{m_{ai},m_s\}$ and $\{m_{c\upsilon},m_t\}$.  }
		\label{tab1}
		\centering
		\begin{tabular}{c||c|c}
			\hline\hline
			Scenario & $\{m_{ai},m_s\}$ & $\{m_{c\upsilon},m_t\}$\\
			\hline\hline
			Scenario $1$ & INT & INT\\
			\hline
			Scenario $2$ & NINT & INT\\
			\hline
			Scenario $3$ & INT & NINT\\
			\hline
			Scenario $4$ & NINT & NINT\\
			\hline
		\end{tabular}\label{tab} 
	\end{table}
	
	Since we have represented the pdf and cdf of SR fading in generalized forms for $\nu\in\{\textmd{INT},\textmd{NINT}\}$, Scenarios $1$ and $2$ in Table \ref{tab} can be equivalently given as Case $1$:\ INT/NINT $\{m_{ai},m_s\}$ and INT $\{m_{c\upsilon},m_t\}$. Similarly, we have Case $2$: INT/NINT $\{m_{ai},m_s\}$ and NINT $\{m_{c\upsilon},m_t\}$. Hence, in what follows, we take into account these two cases for the outage performance analysis of the satellite and IoT networks of the considered OSTN.      
	\section{Statistical Properties of Hybrid Extra-Terrestrial and Terrestrial Interference}\label{statc}
	In this section, we statistically characterize the hybrid interference from ETSs and TSs, i.e., $W_c$, which is given as
	\begin{align}\label{combi}
	W_c=W_s+W_t.
	\end{align}
	
	To proceed, we require the pdf of sum of i.i.d. SR variates (i.e., $W_s$) as well as the sum of i.i.d. Nakagami-\emph{m} variates (i.e., $W_t$). So, we first derive the pdf of $W_s$ as given below.      
	
	\newtheorem{lemma}{Lemma}
	\begin{lemma}\label{ws}
		The pdf of $W_s$ can be expressed as
		\begin{align}\label{wst}
		f_{W_{s}}(w)&=\widetilde{\sum_{(\nu,s)}}\,\frac{\Xi_{(\nu,s)}(M_s)}{\eta^\Lambda_s}w^\Lambda\textmd{e}^{-{\Theta}_{(\nu,s)}w}
		\end{align}
		where $\Xi_{(\nu,s)}(M_s)=\alpha^{M_s}_s\prod_{\kappa=1}^{M_s}\zeta_{(\nu,s)}(i_{\kappa})\prod_{j=1}^{M_s-1}\Phi(\sum_{l=1}^{j}i_{l}+j,i_{j+1}+1)$, $\Lambda=\sum_{\kappa=1}^{M_s}i_{\kappa}+M_s$,
		${\Theta}_{(\nu,s)}=\frac{\beta_{(\nu,s)}}{\eta_s}$, $\underset{(\nu,s)}{\widetilde{\sum}}=\sum_{i_1}^{\varpi_{(\nu,s)}}\dots\sum_{i_{M_s}}^{\varpi_{(\nu,s)}}$ and $\Phi(\cdot,\cdot)$ denotes the Beta function \cite[eq. 8.384.1]{grad}.
		
	\end{lemma}
	
	\begin{IEEEproof}
		Since $W_s=\sum_{j=1}^{M_s}\Lambda_{sj}$, the pdf of $W_s$ can be evaluated via $M_s$-fold statistical convolution of independent pdfs $f_{\Lambda_{sj}}(w)$, $j=1,...,M_s$ as \cite{miridakis2014dual}
		\begin{align}\label{st1}
		f_{W_s}(w) = f_{\Lambda_{s1}}(w)*f_{\Lambda_{s2}}(w)*\dots*f_{\Lambda_{sM_s}}(w),
		\end{align}
		where the symbol ``$*$" stands for convolution. Let us first consider the derivation of pdf $f_{W_s}(w)$ for the case $M_s=2$ which results in $W_s=\Lambda_{s1}+\Lambda_{s2}$. Consequently, we have
		\begin{align}\label{intws}
		f_{W_s}(z)&=\int_{0}^{w}f_{\Lambda_{s1}}(x)f_{\Lambda_{s2}}(w-x)dx.
		\end{align}
		Further, by making use of (\ref{pdfrs}) in (\ref{intws}), one can have
		\begin{align}
		f_{W_s}(w)&=\alpha^2_{s}\sum_{i_1=0}^{\varpi_{(\nu,s)}}\sum_{i_2=0}^{\varpi_{(\nu,s)}}\frac{\zeta_{(\nu,s)}(i_{1})\zeta_{(\nu,s)}(i_{2})}{\eta^{i_1+i_2+2}_s}\textmd{e}^{-{\Theta}_{(\nu,s)}w}\int_{0}^{w}x^{i_1}(w-x)^{i_2}dx,
		\end{align}
		which upon using \cite[eq.(3.191.1)]{grad} yields
		\begin{align}\label{19}
		f_{W_s}(w)&=\alpha^2_{s}\sum_{i_1=0}^{\varpi_{(\nu,s)}}\sum_{i_2=0}^{\varpi_{(\nu,s)}}\frac{\zeta_{(\nu,s)}(i_{1})\zeta_{(\nu,s)}(i_{2})}{\eta^{i_1+i_2+2}_s}\textmd{e}^{-{\Theta}_{(\nu,s)}w}\Phi(i_1+1,i_2+1)w^{i_1+i_2+1}.
		\end{align} 
		Following the identical procedure for $M_s=3$, i.e., $W_s=\Lambda_{s1}+\Lambda_{s2}+\Lambda_{s3}$, the resulting pdf can be calculated successively using the convolution of equivalent pdf of $W_s$ for $M_s=2$ in (\ref{19}) and the pdf in (\ref{pdfrs}) as
		\begin{align}\label{mwc}
		f_{W_s}(w)&=\alpha^3_{s}\sum_{i_1=0}^{\varpi_{(\nu,s)}}\sum_{i_2=0}^{\varpi_{(\nu,s)}}\sum_{i_3=0}^{\varpi_{(\nu,s)}}\frac{\zeta_{(\nu,s)}(i_{1})\zeta_{(\nu,s)}(i_{2})\zeta_{(\nu,s)}(i_{3})}{\eta^{i_1+i_2+i_3+3}_s}\Phi(i_1+1,i_2+1)\\\nonumber
		&\times\Phi(i_1+i_2+2,i_3+1)w^{i_1+i_2+i_3+2}\textmd{e}^{-{\Theta}_{(\nu,s)}w}.
		\end{align}
		Applying the similar procedure for $M_s$ convolutions successively, we can deduce the pdf of $W_s$ as (\ref{wst}).
	\end{IEEEproof}
	
	Further, the pdf of $W_t$ (i.e., sum of i.i.d. and equal power terrestrial Nakagami-\emph{m} interferers) can be given as
	\begin{align}\label{wt}
	f_{W_t}(w)=\left(\frac{m_t}{\Omega_t\eta_t}\right)^{m_tM_t}\frac{w^{m_tM_t-1}}{\Gamma(m_tM_t)}\textmd{e}^{-\frac{m_tw}{\Omega_t\eta_t}}.
	\end{align}
	
	Finally, having the pdfs of $W_s$ and $W_t$, we proceed to determine the pdf of $W_c$ in the next lemma.
	
	\begin{lemma}\label{wc}
		The pdf of $W_c$ can be given as 
		\begin{align}\label{wct}
		f_{W_{s}}(w)\!&=\!\widetilde{\sum_{(\nu,s)}}\,\frac{\Xi_{(\nu,s)}(M_s)}{\eta^\Lambda_s}\left(\frac{m_t}{\Omega_t\eta_t}\right)^{m_tM_t}\frac{\Phi(m_tM_t,\Lambda)}{\Gamma(m_tM_t)}\\\nonumber
		&\times w^{\Lambda+m_tM_t-1}\textmd{e}^{-\frac{m_tw}{\Omega_t \eta_t}} {_1F}_1\left(\Lambda;\Lambda+m_tM_t;-\tilde{\Theta}_{(\nu,s)}w\right),
		\end{align}
		where $\tilde{\Theta}_{(\nu,s)}=\Theta_{(\nu,s)}-\frac{m_{t}}{\Omega_t\eta_t}$.
	\end{lemma}
	
	\begin{IEEEproof}
		Based on (\ref{combi}), the pdf of $W_c$ can be derived as the convolution of two hybrid i.ni.d. interference variables $W_s$ and $W_t$ according to the following expression 
		\begin{align}\label{intwc}
		f_{W_c}(w)&=\int_{0}^{w}f_{W_s}(x)f_{W_t}(w-x)dx.
		\end{align}
		Upon plugging (\ref{wst}) and (\ref{wt}) in (\ref{intwc}), it results 
		\begin{align}\label{inwc}
		f_{W_{s}}(w)&=\widetilde{\sum_{(\nu,s)}}\,\frac{\Xi_{(\nu,s)}(M_s)}{\eta^\Lambda_s}\frac{1}{\Gamma(m_tM_t)}\left(\frac{m_t}{\Omega_t\eta_t}\right)^{m_tM_t}\textmd{e}^{-\frac{m_tw}{\Omega_t \eta_t}}\int_{0}^{w}x^{\Lambda-1}(w-x)^{m_tM_t-1}\textmd{e}^{-\tilde{\Theta}_{(\nu,s)}x}dx.
		\end{align}
		Finally, evaluating the integral in (\ref{inwc}), using \cite[eq. 3.383.1]{grad}, one can attain (\ref{wct}).
	\end{IEEEproof}
	
	\section{Outage Probability of Satellite Network}\label{per}
	In this section, we evaluate the OP and achievable diversity order of the satellite network of considered OSTN under the cases 1 and 2 as discussed previously in Section \ref{srf}.
	
	For a target rate $\mathcal{R}_{p}$, the OP of the primary satellite network with selected IoT network $C_{k^{\ast}}$$-$$D_{k^{\ast}}$ can be determined as
	\begin{align}\label{cmqr}
	\mathcal{P}^{\textmd{sat}}_{\textmd{out}}(\mathcal{R}_{p}) 
	&=\textmd{Pr}\left[\Lambda_{ac_{k^{\ast}}b}<\gamma_{p}\right]=\mathbb{E}\{\left(\textmd{Pr}\left[\Lambda_{ac_{k}b}<\gamma_{p}|W_c=w\right]\right)^K\},
	\end{align}
	where $\gamma_{p}=2^{2\mathcal{R}_{p}}-1$ is a threshold and $\mathbb{E}\{\cdot\}$ is the statistical expectation. Here, we highlight that the OP analysis based on (\ref{cmqr}) using the exact SINR expression in (\ref{snrs}) is analytically intractable. Therefore, we resort to the tight lower bound analysis for OP of satellite network based on an upper bound on exact SINR in (\ref{snrs}). We now proceed with the OP analysis of satellite network for Case 1 in the following subsection.
	\subsection{Case $1$ (INT/NINT $\{m_{ai},m_s\}$ and INT $\{m_{c\upsilon},m_t\}$)} 
	\subsubsection{Lower Bound OP}
	Let $\tilde{\mathcal{P}}^{\textmd{sat}}_{\textmd{out}}(\mathcal{R}_{p})$ represents the tight lower bound on the exact OP  ${\mathcal{P}}^{\textmd{sat}}_{\textmd{out}}(\mathcal{R}_{p})$ in (\ref{cmqr}). Thus, we have the following theorem.   
	\newtheorem{theorem}{Theorem}
	\begin{theorem}\label{tth2}
		The tight lower bound OP of satellite network $\tilde{\mathcal{P}}^{\textmd{sat}}_{\textmd{out}}(\mathcal{R}_{p})$ for Case 1 can be given as
		\begin{align}\label{asl23}
		\tilde{\mathcal{P}}^{\textmd{sat}}_{\textmd{out}}(\mathcal{R}_{p})&=\left\{ \begin{array}{l}
		\Psi(\mathcal{R}_{p}),\textmd{ if } \gamma_{p} < \mu^{\prime}, \\
		1, \textmd{ if } \gamma_{p} \geq\mu^{\prime},
		\end{array}\right.
		\end{align}
		where $\Psi(\mathcal{R}_{p})$ is given by 
			\begin{align}\label{si3}
		\Psi(\mathcal{R}_{p})&=\sum_{n=0}^{K}\binom{K}{n}(-1)^n\alpha^n_{ac}\sum_{S_{(\nu,m)}\in\mathcal{S}}\frac{n!}{\prod_{m=0}^{\varpi_{(\nu,ac)}}s_m!}\prod_{m=0}^{\varpi_{(\nu,ac)}}(\mathcal{A}_{(\nu,m)})^{s_m}\sum_{S_p\in \mathcal{T}_1}\frac{n!}{\prod_{p=0}^{m_{cb}-1}s_p!}\!\!\\\nonumber
		&\times\prod_{p=0}^{m_{cb}-1}(\mathcal{B}_p)^{s_p}\tilde{\gamma}^{\Delta_{(\nu)}}_p\textmd{e}^{-\tilde{\Theta}_{(\nu,ac)} n \tilde{\gamma}_{p}}\sum_{q=0}^{\Delta_{(\nu)}}\binom{\Delta_{(\nu)}}{q}\widetilde{\sum_{(\nu,s)}}\,\frac{\Xi_{(\nu,s)}(M_s)}{\eta^{\Lambda}_s}\left(\frac{m_t}{\Omega_t\eta_t}\right)^{m_t M_t}\frac{\Phi(m_t M_t,\Lambda)}{\Gamma(m_t M_t)}\\\nonumber
		&\times\frac{\Gamma(\tau(q))}{\left(\tilde{\Theta}_{(\nu,ac)}n\tilde{\gamma}_p+\frac{m_t}{\Omega_t \eta_t}\right)^{\tau{(q)}}}{}_2F_1\left(\Lambda,\tau(q);\tau(0);\frac{-\tilde{\Theta}_{(\nu,s)}}{\tilde{\Theta}_{(\nu,ac)}n\tilde{\gamma}_p+\frac{m_t}{\Omega_t \eta_t}}\right),
		\end{align}
		with $\mu^{\prime}=\frac{\mu}{1-\mu}$, $\tilde{\gamma}_{p}=\frac{\gamma_{p}}{\mu-(1-\mu)\gamma_{p}}$,  $\mathcal{S}=\{S_{(\nu,m)}|{\sum}_{m=0}^{\varpi_{(\nu,ac)}}s_{m}=n\}$, $\Delta_{(\nu,ac)}={\sum}_{m=0}^{\varpi_{(\nu,ac)}}m s_{m}$, $\mathcal{A}_{(\nu,m)}={\sum}_{\kappa=m}^{\varpi_{(\nu,ac)}}\frac{\zeta_{(\nu,ac)}(\kappa)}{(\eta_{a})^{\kappa+1}}\frac{\kappa!}{m!} (\Theta_{(\nu,ac)})^{-(\kappa+1-m)}$, $\mathcal{T}_1=\{S_{p}|{\sum}_{p=0}^{m_{cb}-1}s_{p}=n\}$, $\Delta_{cb}={\sum}_{p=0}^{m_{cb}-1}p s_{p}$, $\mathcal{B}_{p}=\frac{1}{p!}(\frac{m_{cb}}{\Omega_{cb}\eta_c})^p$,
		$\tilde\Theta_{(\nu,ac)}=\Theta_{(\nu,ac)}+\frac{m_{cb}}{\Omega_{cb}\eta_{c}}$, $\Theta_{(\nu,ac)}=\frac{\beta_{(\nu,ac)}}{\eta_a}$,  $\Delta_{(\nu)}=\Delta_{(\nu,ac)}+\Delta_{cb}$, $\{s_{m}\}$ and $\{s_{p}\}$ are nonnegative integers along with the function $\tau(x)=\Lambda+m_t M_t +x$, for some $x$.
	\end{theorem}
	\begin{IEEEproof}
		See Appendix \ref{appA}.
	\end{IEEEproof}
	
	\textit{Remark 1:} In (\ref{asl23}), $\gamma_{p} < \mu^{\prime}$ is the necessary condition to allow secondary spectrum access for IoT network, otherwise an outage event is induced making the primary communications unsuccessful. Hence, the maximum rate $\mathcal{R}_{p}$ for satellite network is constrained as $\mathcal{R}_{p} < \frac{1}{2}\log_2(1+\mu^{\prime})$.  
	\subsubsection{Asymptotic OP}
	We derive the asymptotic OP of the satellite network at high SNR (i.e. $(\eta_a, \eta_c)\rightarrow\infty$) to assess its achievable diversity order in the following corollary.
	\newtheorem{corollary}{Corollary}
	\begin{corollary}\label{cor1}
		The asymptotic OP of satellite network for Case 1 under $\gamma_{p} < \mu^{\prime}$ and $\eta_a=\eta_c=\eta$ can be given as
		\begin{align}\label{asyp}
		\tilde{\mathcal{P}}^{\textmd{sat}}_{\textmd{out},\infty}(\mathcal{R}_{p})&=\left\{ \begin{array}{l}
		\left(\frac{\alpha_{ac} \tilde{\gamma}_p}{\eta}\right)^K \psi_1(K), \textmd{ if } m_{cb} > 1, \\
		\left(\frac{\alpha_{ac} \tilde{\gamma}_p}{\eta}+\frac{\tilde{\gamma}_p}{\Omega_{cb}\eta}\right)^K \psi_1(K), \textmd{ if } m_{cb} = 1,
		\end{array}\right.
		\end{align}
		where the function $\psi_1(x)$ is defined as
		\begin{align}
		\!\psi_1(x)&=\!\sum_{q=0}^{x} \binom{x}{q}\widetilde{\sum_{(\nu,s)}}\,\frac{\Xi_{(\nu,s)}(M_s)}{\eta^{\Lambda}_s}\left(\frac{m_t}{\Omega_t\eta_t}\right)^{-(q+\Lambda)}\!\!\\\nonumber
		&\times\frac{\Phi(m_t M_t,\Lambda)}{\Gamma(m_t M_t)}\Gamma{(\tau(q))}{}_2F_1\left(\Lambda,\tau(q) ;\tau(0);\frac{-\tilde{\Theta}_{(\nu,s)}\Omega_t \eta_t}{m_t}\right),
		\end{align}
	\end{corollary}
	\begin{IEEEproof}
		See Appendix \ref{appB}. 
	\end{IEEEproof}
	
	Note that in Corollary \ref{cor1}, the transmit powers $\eta_s$ and $\eta_t$ correspond to interfering ETSs and TSs, respectively, and are kept fixed, i.e., condition $(a)$ is followed. However, for condition $(b)$, when $\eta_s$ and $\eta_t$ vary proportionally with $\eta$, e.g., $\eta_s=\eta_t=\lambda\eta$, for some constant $\lambda$, the asymptotic OP takes on the form as given below.  
	\begin{corollary}\label{corr2}
		The asymptotic OP of satellite network for Case 1 under $\gamma_{p} < \mu^{\prime}$ and $\eta_s=\eta_t=\lambda\eta$ can be given as
		\begin{align}\label{asyp2}
		\tilde{\mathcal{P}}^{\textmd{sat}}_{\textmd{out},\infty}(\mathcal{R}_{p})&=\left\{ \begin{array}{l}
		\left({\alpha_{ac} \tilde{\gamma}_p}\right)^K \psi_2(K), \textmd{ if } m_{cb} > 1,\\
		\left({\alpha_{ac} \tilde{\gamma}_p}\!+\!\frac{\tilde{\gamma}_p}{\Omega_{cb}}\right)^K \psi_2(K),\textmd{ if } m_{cb} = 1,
		\end{array}\right.
		\end{align}
		where the function $\psi_2(x)$ is defined as
		\begin{align}
		\psi_2(x)&=\widetilde{\sum_{(\nu,s)}}\,\Xi_{(\nu,s)}(M_s)\left(\frac{m_t}{\Omega_t}\right)^{-(x+\Lambda)}\frac{\Phi(m_t M_t,\Lambda)}{\Gamma(m_t M_t)}\lambda^x\Gamma{(\tau(x))}{}_2F_1\left(\Lambda,\tau(x);\tau(0);\rho_{(\nu,s)}\right),
		\end{align}
		where $\rho_{(\nu,s)}=1-\frac{\beta_{(\nu,s)}\Omega_t}{m_t}$.
	\end{corollary}
	\begin{IEEEproof}
		Considering $\eta_s=\eta_t=\lambda\eta$ in (\ref{asyp}) and neglecting the higher order terms, one can determine (\ref{asyp2}).
	\end{IEEEproof}
	
	In the next subsection, we proceed for the OP analysis of satellite network under Case $2$. 
	\subsection{Case $2$ (INT/NINT $\{m_{ai},m_s\}$ and NINT $\{m_{c\upsilon},m_t\}$)}
	\subsubsection{Lower Bound OP} In this subsection, we derive the OP of satellite network for Case 2. We have the following theorem.
	\begin{theorem}\label{tth2n}
		The tight lower bound OP of satellite network $\tilde{\mathcal{P}}^{\textmd{sat}}_{\textmd{out}}(\mathcal{R}_{p})$ for Case 2 can be given as
		\begin{align}\label{asl23n}
		\tilde{\mathcal{P}}^{\textmd{sat}}_{\textmd{out}}(\mathcal{R}_{p})&=\left\{ \begin{array}{l}
		\overline{\Psi}(\mathcal{R}_{p}),\textmd{ if } \gamma_{p} < \mu^{\prime} ,\\
		1, \textmd{ if } \gamma_{p} \geq\mu^{\prime},
		\end{array}\right.
		\end{align}
		where $\overline{\Psi}(\mathcal{R}_{p})$ is given by 
		\begin{align}\label{si3n}
		\overline{\Psi}(\mathcal{R}_{p})&=\sum_{n=0}^{K}\binom{K}{n}(-1)^n\alpha^n_{ac}\sum_{S_{(\nu,m)}\in\mathcal{S}}\frac{n!}{\prod_{m=0}^{\varpi_{(\nu,ac)}}s_m!}\prod_{m=0}^{\varpi_{(\nu,ac)}}(\mathcal{A}_{(\nu,m)})^{s_m}\sum_{v=0}^{n}\binom{n}{v}\frac{(-1)^v}{(\Gamma(m_{cb}))^v}
		\\\nonumber
		&\times\sum_{\overline{S}_p\in \overline{\mathcal{T}}}\frac{v!}{\prod_{p=0}^{\infty}\overline{s}_p!}\prod_{p=0}^{\infty}(\overline{\mathcal{B}}_p)^{\overline{s}_p}\tilde{\gamma}^{\overline{\Delta}_{(\nu)}}_p\textmd{e}^{-({\Theta}_{(\nu,ac)} n \tilde{\gamma}_{p})}\widetilde{\sum_{(\nu,s)}}\,\frac{\Xi_{(\nu,s)}(M_s)}{\eta^{\Lambda}_s}\left(\frac{m_t}{\Omega_t\eta_t}\right)^{m_t M_t}\frac{\Phi(m_t M_2,\Lambda)}{\Gamma(m_t M_t)}\\\nonumber
		&\times\sum_{g=0}^{\infty}\frac{(\Lambda)_g (-\tilde{\Theta}_{(\nu,s)})^g}{(\tau(0))_g g!}\Gamma(\tau(g)) U\left(\tau(g),\tau(g+\overline{\Delta}_{(\nu)}+1);\Theta_{(\nu,ac)} n \tilde{\gamma}_p + \frac{m_t}{\Omega_t \eta_t}\right),
		\end{align}
		with $U(\cdot,\cdot;\cdot)$ as the confluent hypergeometric function \cite[eq. 9.211.4]{grad}, $\overline{\mathcal{T}}=\{\overline{S}_{p}|{\sum}_{p=0}^{\infty}\overline{s}_{p}=v\}$, $\overline{\Delta}_{cb}={\sum}_{p=0}^{\infty}(p+m_{cb}) s_{m}$, $\overline{\mathcal{B}}_{p}=\frac{(-1)^p}{p!(m_{cb}+p)}\left(\frac{m_{cb}}{\Omega_{cb}\eta_c}\right)^{p+m_{cb}}$, $\overline{\Delta}_{(\nu)}=\Delta_{(\nu,ac)}+\overline{\Delta}_{cb}$  and $\{\overline{s}_{p}\}$ are nonnegative integers.
		\begin{IEEEproof}
			See Appendix \ref{appC}. 
		\end{IEEEproof}
	\end{theorem}
	\subsubsection{Asymptotic OP} We now derive the asymptotic OP of the satellite network at high SNR for condition $(a)$ below.
	\begin{corollary}\label{cor1n}
		The asymptotic OP of satellite network for Case 2 under $\gamma_{p} < \mu^{\prime}$ and $\eta_a=\eta_c=\eta$ can be given as
		\begin{align}\label{asypn}
		\tilde{\mathcal{P}}^{\textmd{sat}}_{\textmd{out},\infty}(\mathcal{R}_{p})&=\left\{ \begin{array}{l}
		\Big(\frac{\alpha_{ac} \tilde{\gamma}_p}{\eta}\Big)^K\psi_1(K),
		\textmd{ if } m_{cb} > 1, 
		\\\left(\frac{1}{\Gamma(m_{cb}+1)}\right)^K\left(\frac{m_{cb}\tilde{\gamma}_p}{\Omega_{cb}\eta}\right)^{m_{cb}K}\overline{\psi}_1(m_{cb}K),\textmd{ if } m_{cb} < 1,
		\end{array}\right.
		\end{align}
		where the function $\overline{\psi}_1(x)$ is defined as
		\begin{align}\label{34}
		\overline{\psi}_1(x)&=\widetilde{\sum_{(\nu,s)}}\,\frac{\Xi_{(\nu,s)}(M_s)}{\eta^{\Lambda}_s}\left(\frac{m_t}{\Omega_t\eta_t}\right)^{m_t M_t}\frac{\Phi(m_t M_t,\Lambda)}{\Gamma(m_t M_t)}\\\nonumber
		&\times\sum_{g=0}^{\infty}\frac{(\Lambda)_g (-\tilde{\Theta}_{(\nu,s)})^g}{(\tau(0))_g g!}\Gamma{(\tau(g))}U\left(\tau(g);\tau(g+x+1);\frac{m_t}{\Omega_t \eta_t}\right).
		\end{align}	
	\end{corollary}
	\begin{IEEEproof}
		See Appendix \ref{appD}.
	\end{IEEEproof}
	
		For condition $(b)$, the asymptotic OP is given as follows.
	\begin{corollary}\label{cor1n1}
		The asymptotic OP of satellite network for Case 2 under $\gamma_{p} < \mu^{\prime}$ and $\eta_s=\eta_t=\lambda\eta$ can be given as
		\begin{align}\label{asypn1}
		\tilde{\mathcal{P}}^{\textmd{sat}}_{\textmd{out},\infty}(\mathcal{R}_{p})&=\left\{ \begin{array}{l}
		\left(\alpha_{ac} \tilde{\gamma}_p\right)^K\psi_2(K),\textmd{ if } m_{cb} > 1, 
		\\\left(\frac{1}{\Gamma(m_{cb}+1)}\right)^K\left(\frac{m_{cb}\tilde{\gamma}_p}{\Omega_{cb}}\right)^{m_{cb}K}\overline{\psi}_2(m_{cb}K),\textmd{ if } m_{cb} < 1,
		\end{array}\right.
		\end{align}
		where the function $\overline{\psi}_2(x)$ is defined as
		\begin{align}
		\overline{\psi}_2(x)&=\widetilde{\sum_{(\nu,s)}}\Xi_{(\nu,s)}(M_s)\left(\frac{m_t}{\Omega_t}\right)^{-(\Lambda+x)}\frac{\Phi(m_t M_t,\Lambda)}{\Gamma(m_t M_t)}\sum_{g=0}^{\infty}\frac{(\Lambda)_g (\rho_{(\nu,s)})^g}{(\tau(0))_g g!}\lambda^x\Gamma(\tau(g+x)).
		\end{align}
	\end{corollary}
	\begin{IEEEproof}
			For $m_{cb}>1$, the proof is the same as in Corollary \ref{corr2}. However, for $m_{cb}<1$, we need to simplify first the function $U(\cdot,\cdot;\cdot)$ present in $\overline{\psi}_1(x)$. Hereby, we invoke the series expansion $U(a,b;z)=\frac{\Gamma(b-1)}{\Gamma(a)}z^{1-b}+O(z^{b-2})$, for $b>2$ under small $z$ \cite[eq. 13.5.6]{abm},  where $a$, $b$ are the constants and $O(\cdot)$ represents the higher order terms. First, applying this for $U(\cdot,\cdot;\cdot)$ in $\overline{\psi}_1(x)$ in (\ref{34}) along with $\eta_s=\eta_t=\lambda\eta$ and then substituting the result in (\ref{asypn}), one can attain (\ref{asypn1}).
	\end{IEEEproof}

	\emph{Remark 2:} For Case 1, the asymptotic OP of satellite network is proportional to $\eta^{-K}$ under condition $(a)$ as seen from (\ref{asyp}). Irrespective of the INT/NINT values of Nakagami-m based SR fading, the diversity order remains $K$ for INT values of $m_{cb}$. On the contrary, the diversity order of satellite network for Case 2 (i.e., NINT $m_{cb}$) is $m_{cb}K$ for $m_{cb}<1$ and $K$ for $m_{cb}>1$ according to the asymptotic OP expression (\ref{asypn}). However, for condition $(b)$, the diversity order reduces to zero as seen from (\ref{asyp2}) and (\ref{asypn1}) for Cases 1 and 2, respectively.
	
	\section{Outage Probability of IoT Network}\label{sec}
	As done for satellite network, in this section, we evaluate the OP and achievable diversity order of the IoT network of considered OSTN under the cases $1$ and $2$.
	
	For a target rate $\mathcal{R}_{\textmd{S}}$, based on the SINR in (\ref{secb}), the OP of the selected secondary IoT network $C_{k^{\ast}}$$-$$D_{k^{\ast}}$ can be given as
	\begin{align}\label{cm0sec}
	&\mathcal{P}^\textmd{IoT}_{\textmd{out}}(\mathcal{R}_{\textmd{S}}) =\textmd{Pr}[{\Lambda}_{a{c_{k^\ast}}{d_{k^\ast}}}<\gamma_s]{=}\mathbb{E}\left\{\textmd{Pr}\left[\frac{\mu\hat{\Lambda}_{c_{k^{\ast}}d_{k^{\ast}}}(\hat{\Lambda}_{ac_{k^{\ast}}}+1)}{\mu\hat{\Lambda}_{c_{k^{\ast}}d_{k^{\ast}}}+\hat{\Lambda}_{ac_{k^{\ast}}}+1}
	<\mu^{\prime}\gamma_{s}\Big|W_c=w\right]\right\},
	\end{align}
	where $\gamma_{s}=2^{2\mathcal{R}_{\textmd{S}}}-1$ is a threshold. Note that the exact OP analysis according to (\ref{cm0sec}) is analytically intractable due to the two factors, i.e., the involvement of too many random variables in SINR ${\Lambda}_{a{c_{k^\ast}}{d_{k^\ast}}}$ and the statistical characterization of ${\Lambda}_{a{c_{k^\ast}}{d_{k^\ast}}}$ for the selected IoT network $C_{k^{\ast}}$$-$$D_{k^{\ast}}$. Here, we encompass the first factor by applying the bound $\frac{XY}{X+Y}\leq\min(X,Y)$ to evaluate a lower bound on exact OP in (\ref{cm0sec}) (say $\tilde{\mathcal{P}}^\textmd{IoT}_{\textmd{out}}(\mathcal{R}_{\textmd{S}})$) as
	\begin{align}\label{cm0sec1}
	\tilde{\mathcal{P}}^\textmd{IoT}_{\textmd{out}}(\mathcal{R}_{\textmd{S}}) &=\mathbb{E}\{\tilde{\mathcal{P}}^\textmd{IoT}_{\textmd{out}}(\mathcal{R}_{\textmd{S}}|W_c=w)\},
	\end{align}
	where the conditional OP $\tilde{\mathcal{P}}^\textmd{IoT}_{\textmd{out}}(\mathcal{R}_{\textmd{S}}|w)$ can be expressed as 
	\begin{align}\label{39}
	\tilde{\mathcal{P}}^\textmd{IoT}_{\textmd{out}}(\mathcal{R}_{\textmd{S}}|w)&=\textmd{Pr}[\min{(\mu\hat{\Lambda}_{c_{k^{\ast}}d_{k^{\ast}}},\hat{\Lambda}_{ac_{k^{\ast}}}\!\!+\!1)}
	\!<{\mu^{\prime}\gamma_{s}}|w].
	\end{align}
	After a variable transformation for $\hat{\Lambda}_{ac_{k^{\ast}}}+1$ and some manipulation, $\tilde{\mathcal{P}}^\textmd{IoT}_{\textmd{out}}(\mathcal{R}_{\textmd{S}}|w)$ in (\ref{39}) can be expressed as
	\begin{align}\label{aql23}
	\tilde{\mathcal{P}}^\textmd{IoT}_{\textmd{out}}(\mathcal{R}_{\textmd{S}}|w)\!&=\!\left\{ \begin{array}{l}
	F_{\mu\hat{\Lambda}_{c_{k^{\ast}}d_{k^{\ast}}}}(\mu^{\prime}\gamma_{s}|w),\textmd{ if } \gamma_{s} < \frac{1}{\mu^{\prime}}, \\
	{F}_{\mu\hat{\Lambda}_{c_{k^{\ast}}d_{k^{\ast}}}}(\mu^{\prime}\gamma_{s}|w)
	+{F}_{\hat{\Lambda}_{ac_{k^{\ast}}}}({\mu^{\prime}\gamma_{s}}-1|w)\overline{F}_{\mu\hat{\Lambda}_{c_{k^{\ast}}d_{k^{\ast}}}}(\mu^{\prime}\gamma_{s}|w), \textmd{ if } \gamma_{s} \geq \frac{1}{\mu^{\prime}},
	\end{array}\right.
	\end{align}
	where $\overline{F}_{X}(\cdot|w)=1-F_{X}(\cdot|w)$. In the next subsection, we evaluate the OP of IoT network based on (\ref{cm0sec1}) for Case 1.

	\subsection{Case $1$ (INT/NINT $\{m_{ai},m_s\}$ and INT $\{m_{c\upsilon},m_t\}$)} 
	\subsubsection{Lower Bound OP} As can be observed, in (\ref{aql23}), we first need to obtain the cdf ${F}_{\hat{\Lambda}_{ac_{k^{\ast}}}}(x|w)$ for selected IoT network according to the following lemma. 
	\begin{lemma}\label{lem1}
		The cdf ${F}_{\hat{\Lambda}_{ac_{k^{\ast}}}}(x|w)$ for the selected IoT network $C_{k^{\ast}}$$-$$D_{k^{\ast}}$ under Case $1$ is given by 
		\begin{align}\label{fac}
		{F}_{\hat{\Lambda}_{ac_{k^{\ast}}}}(x|w)&=\varphi_1(x|w)+\varphi_2(x|w),
		\end{align}
	\end{lemma}
	where
	\begin{align}\label{psi1}
	\varphi_1(x|w)&=\frac{K}{\Gamma(m_{cb})}\bigg(\frac{m_{cb}}{\Omega_{cb}\eta_c}\bigg)^{m_{cb}}\sum_{n=0}^{K-1}
	\binom{K-1}{n}(-1)^{n}\alpha^{n+1}_{ac}\sum_{S_{(\nu,m)}\in\mathcal{S}}\frac{n!}{\prod_{m=0}^{\varpi_{(\nu,ac)}}s_m!}\\\nonumber
	&\times\prod_{m=0}^{\varpi_{(\nu,ac)}}(\mathcal{A}_{(\nu,m)})^{s_m}\sum_{S_p\in \mathcal{T}_1}\frac{n!}{\prod_{p=0}^{m_{cb}-1}s_p!}\prod_{p=0}^{m_{cb}-1}(\mathcal{B}_p)^{s_p}\frac{\Gamma(\Delta_{(\nu)}+m_{cb})}
	{\vartheta^{\Delta_{(\nu)}+m_{cb}}_{(\nu,n)}}
	\sum_{\kappa=0}^{\varpi_{(\nu,ac)}}\frac{\zeta_{(\nu,ac)}(\kappa)}{\eta^{\kappa+1}_{a}} \\\nonumber
	&\times\left[\frac{\Upsilon(\kappa+1,\Theta_{(\nu,ac)}x(w+1))}
	{\Theta_{(\nu,ac)}^{\kappa+1}}\right.\left.-\sum_{q=0}^{\Delta_{(\nu)}+m_{cb}-1}
	\frac{\left(\vartheta_{(\nu,n)}\right)^{q}}{q!\omega^{\kappa+q+1}_{(\nu,n)}}{\Upsilon\left(\kappa+q+1,\omega_{(\nu,n)}x(w+1)\right)}
	\right],
	\end{align}
	and
	\begin{align}\label{psi2}
	\varphi_2(x|w)&=K\sum_{\kappa=0}^{\varpi_{(\nu,ac)}}\frac{\zeta_{(\nu,ac)}(\kappa)}{\eta^{\kappa+1}_{a}}\sum_{n=0}^{K-1}\binom{K-1}{n}(-1)^{n}\alpha^{n+1}_{ac}\sum_{S_{(\nu,m)}\in\mathcal{S}}\frac{n!}{\prod_{m=0}^{\varpi_{(\nu,ac)}}s_m!}\\\nonumber
	&\times\prod_{m=0}^{\varpi_{(\nu,ac)}}(\mathcal{A}_{(\nu,m)})^{s_m}\sum_{S^{\prime}_p\in \mathcal{T}_2}\frac{(n+1)!}{\prod_{p=0}^{m_{cb}-1}s^{\prime}_p!}\prod_{p=0}^{m_{cb}-1}(\mathcal{B}_p)^{s^{\prime}_p}\frac{\Upsilon\left(\kappa+\Delta^{\prime}_{(\nu)}+1,\omega_{(\nu,n)}x(w+1)\right)}{\omega^{\kappa+\Delta^{\prime}_{(\nu)}+1}_{(\nu,n)}},
	\end{align}
	with $\vartheta_{(\nu,n)}=n\Theta_{(\nu,ac)}+\frac{(n+1)m_{cb}}{\Omega_{cb}\eta_c}$ and $\omega_{(\nu,n)}=(n+1)\tilde{\Theta}_{(\nu,ac)}$.
	\begin{IEEEproof}
		See Appendix \ref{appE}.
	\end{IEEEproof}
	Further, by realizing the fact that the selection strategy in (\ref{ous}) is independent of the $C_k-D_k$ link, the cdf $F_{\mu\hat{\Lambda}_{c_{k^{\ast}}d_{k^{\ast}}}}(x|w)$ in (\ref{aql23}) does not follow the order statistics, i.e., $F_{\mu\hat{\Lambda}_{c_{k^{\ast}}d_{k^{\ast}}}}(x|w)=F_{\mu\hat{\Lambda}_{c_{k}d_{k}}}(x|w)$. Hence, we have
	\begin{align}\label{fsac}
	F_{\mu\hat{\Lambda}_{c_{k^{\ast}}d_{k^{\ast}}}}(x|w)&=\frac{1}{\Gamma(m_{cd})}\Upsilon \left(m_{cd}, \frac{m_{cd} x(w+1)}{\Omega_{cd}\eta_{c}\mu}\right).
	\end{align}
	Having derived the required cdfs ${F}_{\hat{\Lambda}_{ac_{k^{\ast}}}}(x|w)$ and $F_{\mu\hat{\Lambda}_{c_{k^{\ast}}d_{k^{\ast}}}}(x|w)$, we calculate the OP of secondary IoT network in the following theorem. 
	\begin{theorem}\label{th1}
		The lower bound OP of the secondary IoT network $\tilde{\mathcal{P}}^{\textmd{IoT}}_{\textmd{out}}(\mathcal{R}_{s})$ for Case $1$ is given as
		\begin{align}\label{ajsl23}
		\tilde{\mathcal{P}}^{\textmd{IoT}}_{\textmd{out}}(\mathcal{R}_{s})&=\left\{ \begin{array}{l}
		\Psi_1(\mathcal{R}_{s}),\textmd{ if } \gamma_{s} < \frac{1}{\mu^{\prime}}, \\
		\Psi_1(\mathcal{R}_{s})+\Psi_2(\mathcal{R}_{s})+\Psi_3(\mathcal{R}_{s}),\textmd{ if } \gamma_{s} \geq\frac{1}{\mu^{\prime}},
		\end{array}\right.
		\end{align}
		where $\Psi_1(\mathcal{R}_{s})$ and $\Psi_2(\mathcal{R}_{s})$ are given as 
		\begin{align}\label{si11}
		\Psi_1(\mathcal{R}_{s})&=1-\sum_{l=0}^{m_{cd}-1}\frac{1}{l!}\left(\frac{m_{cd}\gamma_s}{\Omega_{cd}\eta_c(1-\mu)}\right)^l \textmd{e}^{\frac{-m_{cd}\gamma_s}{\Omega_{cd}\eta_c(1-\mu)}}\sum_{q=0}^{l}\binom{l}{q}\widetilde{\sum_{(\nu,s)}}\,\frac{\Xi_{(\nu,s)}(M_s)}{\eta^{\Lambda}_s}\left(\frac{m_t}{\Omega_t\eta_t}\right)^{m_t M_t}\\\nonumber
		&\times\frac{\Phi(m_t M_t,\Lambda)}{\Gamma(m_t M_t)}\frac{\Gamma(\tau(q))}{\chi^{\tau(q)}_{c,t}}{}_2F_1\left(\Lambda,\tau(q);\tau(0);\frac{-\tilde{\Theta}_{(\nu,s)}}{\chi_{c,t}}\right)
		\end{align}
		and 
		\begin{align}\label{si22}
		&\Psi_2(\mathcal{R}_{s})=K\!\!\sum_{\kappa=0}^{\varpi_{(\nu,ac)}}\!\!\frac{\zeta_{(\nu,ac)}(\kappa)}{\eta^{\kappa+1}_{a}}\sum_{n=0}^{K-1}
		\binom{K-1}{n}(-1)^{n}\alpha^{n+1}_{ac}\sum_{S_{(\nu,m)}\in\mathcal{S}}\frac{n!}{\prod_{m=0}^{\varpi_{(\nu,ac)}}s_m!}\prod_{m=0}^{\varpi_{(\nu,ac)}}(\mathcal{A}_{(\nu,m)})^{s_m}\\\nonumber
		&\times\sum_{S^{\prime}_p\in \mathcal{T}_2}\frac{(n+1)!}{\prod_{p=0}^{m_{cb}-1}s^{\prime}_p!}\prod_{p=0}^{m_{cb}-1}(\mathcal{B}_p)^{s^{\prime}_p}\frac{\Gamma(\kappa+\Delta^{\prime}_{(\nu)}+1)}{\omega^{\kappa+\Delta^{\prime}_{(\nu)}+1}_{(\nu,n)}}\sum_{l=0}^{m_{cd}-1}\frac{1}{l!}\left(\frac{m_{cd}\gamma_s}{\Omega_{cd}\eta_c(1-\mu)}\right)^l \textmd{e}^{\frac{-m_{cd}\gamma_s}{\Omega_{cd}\eta_c(1-\mu)}}\\\nonumber
		&\times\sum_{q=0}^{l}\binom{l}{q}\widetilde{\sum_{(\nu,s)}}\,\frac{\Xi_{(\nu,s)}(M_s)}{\eta^{\Lambda}_s}\left(\frac{m_t}{\Omega_t\eta_t}\right)^{m_t M_t}\frac{\Phi(m_t M_t,\Lambda)}{\Gamma(m_t M_t)}\left[\frac{\Gamma(\tau(q))}{\chi^{\tau(q)}_{c,t}}{}_2F_1\bigg(\Lambda,\tau(q);\tau(0);\frac{-\tilde{\Theta}_{(\nu,s)}}{\chi_{c,t}}\bigg)\right.\\\nonumber
		&-\sum_{v=0}^{\kappa+\Delta^{\prime}_{(\nu)}}\frac{(\omega_{(\nu,n)}\tilde{\gamma}_s)^v}{v!}\textmd{e}^{-\omega_{(\nu,n)}\tilde{\gamma}_s}\sum_{g=0}^{v}\binom{v}{g}\left.\frac{\Gamma(\tau(q+g))}{(\omega_{(\nu,n)}\tilde{\gamma}_s+\chi_{c,t})^{\tau(q+g)}} {}_2F_1\left(\Lambda,\tau(q+g);\tau(0);\frac{-\tilde{\Theta}_{(\nu,s)}}{\omega_{(\nu,n)}\tilde{\gamma}_s+\chi_{c,t}}\right)\right],
		\end{align}
		respectively. 

		Also, $\Psi_3(\mathcal{R}_{s})$ is given by
		\begin{align}\label{si33}
		\Psi_3(\mathcal{R}_{s})&=\frac{K}{\Gamma(m_{cb})}\bigg(\frac{m_{cb}}{\Omega_{cb}\eta_c}\bigg)^{m_{cb}}\sum_{n=0}^{K-1}
		\binom{K-1}{n}(-1)^{n}\alpha^{n+1}_{ac}\sum_{S_{(\nu,m)}\in\mathcal{S}}\frac{n!}{\prod_{m=0}^{\varpi_{(\nu,ac)}}s_m!}\\\nonumber
		&\times\prod_{m=0}^{\varpi_{(\nu,ac)}}(\mathcal{A}_{(\nu,m)})^{s_m}\sum_{S_p\in \mathcal{T}_1}\frac{n!}{\prod_{p=0}^{m_{cb}-1}s_p!}\prod_{p=0}^{m_{cb}-1}(\mathcal{B}_p)^{s_p}\frac{\Gamma(\Delta_{(\nu)}+m_{cb})}
		{\vartheta^{\Delta_{(\nu)}+m_{cb}}_{(\nu,n)}}\sum_{\kappa=0}^{\varpi_{(\nu,ac)}}\frac{\zeta_{(\nu,ac)}(\kappa)}{\eta^{\kappa+1}_{a}}\\\nonumber
		&\times\sum_{l=0}^{m_{cd}-1}\frac{1}{l!}\left(\frac{m_{cd}\gamma_s}{\Omega_{cd}\eta_c(1-\mu)}\right)^l \textmd{e}^{\frac{-m_{cd}\gamma_s}{\Omega_{cd}\eta_c(1-\mu)}}\sum_{d=0}^{l}\binom{l}{d}\widetilde{\sum_{(\nu,s)}}\,\frac{\Xi_{(\nu,s)}(M_s)}{\eta^{\Lambda}_s}\left(\frac{m_t}{\Omega_t\eta_t}\right)^{m_t M_t}\frac{\Phi(m_t M_t,\Lambda)}{\Gamma(m_t M_t)}\\\nonumber
		&\times(\Psi_4(\mathcal{R}_{s})-\Psi_5(\mathcal{R}_{s})),
		\end{align}
		where $\Psi_4(\mathcal{R}_{s})$ and $\Psi_5(\mathcal{R}_{s})$ are given by 
		\begin{align}\label{si44}
		\Psi_4(\mathcal{R}_{s})&=\frac{\Gamma(\kappa+1)}{\Theta^{\kappa+1}_{(\nu,ac)}}\left[\frac{\Gamma(\tau(q))}{\chi^{\tau(q)}_{c,t}}{}_2F_1\left(\Lambda,\tau(q);\tau(0);\frac{-\tilde{\Theta}_{(\nu,s)}}{\chi_{c,t}}\right)\right.-\sum_{v=0}^{\kappa}\frac{(\Theta_{(\nu,ac)} \tilde{\gamma}_s)^v}{v!}\textmd{e}^{-\Theta_{(\nu,ac)}\tilde{\gamma}_s}\\\nonumber
		&\left.\sum_{g=0}^{v}\binom{v}{g}\frac{\Gamma(\tau(q+g))}{(\Theta_{(\nu,ac)}\tilde{\gamma}_s+\chi_{c,t})^{\tau(q+g)}}{}_2F_1\left(\Lambda,\tau(q+g);\tau(0);\frac{-\tilde{\Theta}_{(\nu,s)}}{\tilde{\Theta}_{(\nu,ac)}\tilde{\gamma}_s+\chi_{c,t}}\right)\right]
		\end{align}
		and 
\begin{align}\label{si55}
		&\Psi_5(\mathcal{R}_{s})=\sum_{q=0}^{\Delta_{\nu,ac}+m_{cb}-1}\frac{\vartheta^q_{(\nu,n)}}{q!}\frac{\Gamma(\kappa+q+1)}{\omega^{\kappa+q+1}_{(\nu,n)}}\left[\frac{\Gamma(\tau(d))}{\chi^{\tau(d)}_{c,t}}{}_2F_1\left(\Lambda,\tau(d);\tau(0);\frac{-\tilde{\Theta}_{(\nu,s)}}{\chi_{c,t}}\right)\right.\\\nonumber
		&\left.-\sum_{u=0}^{\kappa+q}\frac{(\omega_n \tilde{\gamma}_s)^u}{u!}\textmd{e}^{-\omega_n\tilde{\gamma}_s}\sum_{j=0}^{u}\binom{u}{j}\frac{\Gamma(\tau(d+j))}{(\omega_{(\nu,n)}\tilde{\gamma}_s+\chi_{c,t})^{\tau(d+j)}} {}_2F_1\left(\Lambda,\tau(d+j);\tau(0);\frac{-\tilde{\Theta}_{(\nu,s)}}{\omega_{(\nu,n)}\tilde{\gamma}_s+\chi_{c,t}}\right)\right],
\end{align}
respectively. Herein, various terms are $\tilde{\gamma}_{s}={\mu^{\prime}\gamma_{s}-1}$, $\chi_{c,t}=\frac{m_t}{\Omega_t\eta_t}+\frac{m_{cd}\gamma_s}{\Omega_{cd}\eta_c(1-\mu)}$, $\Delta^{\prime}_{(\nu)}=\Delta_{(\nu,ac)}+\Delta^{\prime}_{cb}$ with all other terms the same as defined previously.

		\begin{figure*}[!t]
			
			\hrule
		\end{figure*}
	\end{theorem}
	\begin{IEEEproof}
		By first making use of series representation \cite[eq. 8.352.6]{grad} for function $\Upsilon(\cdot,\cdot)$ in (\ref{psi1}), (\ref{psi2}) and (\ref{fsac}) and eventually, evaluating (\ref{cm0sec1}) results in (\ref{ajsl23}), where
		\begin{align}
		\Psi_1(\mathcal{R}_{s})=\mathbb{E}(F_{\mu\hat{\Lambda}_{c_{k^{\ast}}d_{k^{\ast}}}}(\mu^{\prime}\gamma_{s}|w)),
		\end{align}
		\begin{align}
		\Psi_2(\mathcal{R}_{s})=\mathbb{E}(\varphi_2({\mu^{\prime}\gamma_{s}}-1|w)\overline{F}_{\mu\hat{\Lambda}_{c_{k^{\ast}}d_{k^{\ast}}}}(\mu^{\prime}\gamma_{s}|w)),
		\end{align}
		and
		\begin{align}
		\Psi_3(\mathcal{R}_{s})=\mathbb{E}(\varphi_1({\mu^{\prime}\gamma_{s}}-1|w)\overline{F}_{\mu\hat{\Lambda}_{c_{k^{\ast}}d_{k^{\ast}}}}(\mu^{\prime}\gamma_{s}|w)).
		\end{align}
		We can compute now $\Psi_1(\mathcal{R}_{s})$, $\Psi_2(\mathcal{R}_{s})$, and $\Psi_3(\mathcal{R}_{s})$ similar to  $\mathcal{I}_1$ in Appendix \ref{appA}.
	\end{IEEEproof}
	
	\subsubsection{Asymptotic OP}  We examine the asymptotic OP of IoT network for its achievable diversity order under condition $(a)$.
	\begin{corollary}\label{cor2}
		The asymptotic OP of IoT network for Case $1$ under $\eta_a=\eta_c=\eta$ is given by 
		\begin{align}\label{asyy2}
			\tilde{\mathcal{P}}^{\textmd{IoT}}_{\textmd{out},\infty}(\mathcal{R}_{s})\!&\simeq\!\left\{ \begin{array}{l}
			\frac{1}{\Gamma(m_{cd}+1)}\left(\frac{m_{cd}\gamma_{s}}{\Omega_{cd}\eta(1-\mu)}\right)^{m_{cd}}\psi_1(m_{cd}), \textmd{ if } \gamma_{s} < \frac{1}{\mu^{\prime}} \textmd{ and } m_{cb} \geq 1,
			\\\frac{1}{\Gamma(m_{cd}+1)}\left(\frac{m_{cd}\gamma_{s}}{\Omega_{cd}\eta(1-\mu)}\right)^{m_{cd}}\psi_1(m_{cd})
			+\left(1\!+\!\frac{1}{\alpha_{ac}\Omega_{cb}}\right)^{K\!-\!1}\left(\frac{\alpha_{ac}\tilde{\gamma}_s}{\eta}\right)^K\psi_1(K),\\ \textmd{ if } \gamma_{s} \geq \frac{1}{\mu^{\prime}} \textmd{ and } m_{cb} =1,
			\\\frac{1}{\Gamma(m_{cd}+1)}\left(\frac{m_{cd}\gamma_{s}}{\Omega_{cd}\eta(1-\mu)}\right)^{m_{cd}}\psi_1(m_{cd})+\left(\frac{\alpha_{ac}\tilde{\gamma}_s}{\eta}\right)^K \psi_1(K), \textmd{ if } \gamma_{s} \geq \frac{1}{\mu^{\prime}} \textmd{ and } m_{cb} >1,
			\end{array}\right.
			\end{align}
		where the function $\psi_1(x)$ is the same as defined in Corollary \ref{cor1}. 
	\end{corollary}
	\begin{IEEEproof}
		The proof follows the Appendix \ref{appE} whereby we approximate the cdf $F_{\mu\hat{\Lambda}_{c_{k^{\ast}}d_{k^{\ast}}}}(x|w) \simeq\frac{1}{\Gamma(m_{cd}+1)}\big(\frac{m_{cd}(w+1)x}{\Omega_{cd}\eta\mu}\big)^{m_{cd}}$ at high SNR. Further, we approximate $F_{\hat{\Lambda}_{ac_{k^{\ast}}}}(x|w)$ by simplifying the term $\left[F_{\digamma_k}(z|w)\right]^{K-1}$ in (\ref{mult}) at high SNR as
		\begin{align}\label{multh}
		\left[F_{\digamma_k}(z|w)\right]^{K-1}\simeq\left[{F}_{\hat{\Lambda}_{ac_k}}(z|w)+{F}_{\hat{\Lambda}_{c_kb}}(z|w)\right]^{K-1},
		\end{align}
		where the product of cdfs leading to higher order is neglected. Upon inserting these cdfs in (\ref{fin}) along with the high SNR approximations of cdfs ${F}_{\hat{\Lambda}_{ac_k}}(z|w)$ and ${F}_{\hat{\Lambda}_{c_kb}}(z|w)$ from Appendix \ref{appB}, one can realize that $\varphi_1(x|w)$ results in higher order. Hence, at high SNR, the cdf  $F_{\hat{\Lambda}_{ac_{k^{\ast}}}}(x|w)$ is dominated by the term $\varphi_2(x|w)$ only which on evaluation yields  
		\begin{align}\label{stgh}
		F_{\hat{\Lambda}_{ac_{k^{\ast}}}}(x|w)\!&\simeq\!\left\{ \begin{array}{l}
		\!\!\!\left(\frac{\alpha_{ac} x }{\eta_c}\right)^K (w+1)^K, \textmd{ if } m_{cb} > 1, \\
		\!\!\!\frac{\alpha_{ac}}{\eta}\left(\frac{\alpha_{ac}}{\eta}\!+\!\frac{1}{\Omega_{cb}\eta}\right)^{K-1} x^K (w\!+\!1)^K, \textmd{ if } m_{cb} = 1.
		\end{array}\right.
		\end{align}
		Having these cdfs required in (\ref{aql23}), one can evaluate (\ref{cm0sec1}) by taking the final expectation as in Appendix \ref{appA} to get (\ref{asyy2}).
	\end{IEEEproof}
	
The asymptotic OP of IoT network under condition $(b)$ is given below.		
	\begin{corollary}\label{cor22}
		The asymptotic OP of IoT network for Case 1 under $\eta_s=\eta_t=\lambda\eta$ is given by 		
			\begin{align}\label{asy2}
			\tilde{\mathcal{P}}^{\textmd{IoT}}_{\textmd{out},\infty}(\mathcal{R}_{s})&\simeq\left\{ \begin{array}{l}
			\frac{1}{\Gamma(m_{cd}+1)}\left(\frac{m_{cd}\gamma_{s}}{\Omega_{cd}(1-\mu)}\right)^{m_{cd}}\psi_2(m_{cd}), \textmd{ if } \gamma_{s} < \frac{1}{\mu^{\prime}} \textmd{ and } m_{cb} \geq 1,
			\\\frac{1}{\Gamma(m_{cd}+1)}\left(\frac{m_{cd}\gamma_{s}}{\Omega_{cd}(1-\mu)}\right)^{m_{cd}}\psi_2(m_{cd})
			+\left(1\!+\!\frac{1}{\alpha_{ac}\Omega_{cb}}\right)^{K\!-\!1}\left(\alpha_{ac}\tilde{\gamma}_s\right)^K\psi_2(K),\\ \textmd{ if } \gamma_{s} \geq \frac{1}{\mu^{\prime}} \textmd{ and } m_{cb} =1,
			\\\frac{1}{\Gamma(m_{cd}+1)}\left(\frac{m_{cd}\gamma_{s}}{\Omega_{cd}(1-\mu)}\right)^{m_{cd}}\psi_2(m_{cd})+\left(\alpha_{ac}\tilde{\gamma}_s\right)^K \psi_2(K), \textmd{ if } \gamma_{s} \geq \frac{1}{\mu^{\prime}} \textmd{ and } m_{cb} >1,
			\end{array}\right.
			\end{align}
		where the function $\psi_2(x)$ is the same as defined in Corollary \ref{corr2}. 
	\end{corollary}
	\begin{IEEEproof}
		The proof is similar to that of Corollary \ref{corr2}. 
	\end{IEEEproof}
	
	
	\subsection{Case $2$ (INT/NINT $\{m_{ai},m_s\}$ and NINT $\{m_{c\upsilon},m_t\}$)}
	\subsubsection{Lower Bound OP} We hereby proceed to derive the OP of IoT network for Case $2$. Similar to that in Case 1, here we need the cdf ${F}_{\hat{\Lambda}_{ac_{k^{\ast}}}}(x|w)$ for selected IoT network as derived in the following lemma. 
	\begin{lemma}\label{lemn1}
		The cdf ${F}_{\hat{\Lambda}_{ac_{k^{\ast}}}}(x|w)$ for the selected IoT network $C_{k^{\ast}}$$-$$D_{k^{\ast}}$ under Case 2 is given by 
		\begin{align}\label{fac2}
		{F}_{\hat{\Lambda}_{ac_{k^{\ast}}}}(x|w)&=\overline{\varphi}_1(x|w)+\overline{\varphi}_2(x|w),
		\end{align}
	\end{lemma}
	where
	\begin{align}\label{psin1}
	\overline{\varphi}_1(x|w)&=\frac{K}{\Gamma(m_{cb})}\bigg(\frac{m_{cb}}{\Omega_{cb}\eta_c}\bigg)^{m_{cb}}\sum_{n=0}^{K-1}
	\binom{K-1}{n}(-1)^{n}\alpha^{n+1}_{ac}\sum_{S_{(\nu,m)}\in\mathcal{S}}\frac{n!}{\prod_{m=0}^{\varpi_{(\nu,ac)}}s_m!}\\\nonumber
	&\times\prod_{m=0}^{\varpi_{(\nu,ac)}}(\mathcal{A}_{(\nu,m)})^{s_m}\sum_{v=0}^{n}\binom{n}{v}\frac{(-1)^v}{(\Gamma(m_{cb}))^v}\sum_{\overline{S}_p\in \overline{\mathcal{T}}}\frac{v!}{\prod_{p=0}^{\infty}\overline{s}_p!}\prod_{p=0}^{\infty}(\overline{\mathcal{B}}_p)^{\overline{s}_p}\sum_{\kappa=0}^{\varpi_{(\nu,ac)}}\frac{\zeta_{(\nu,ac)}(\kappa)}{\eta^{\kappa+1}_{a}}
	\\\nonumber
	&\times\sum_{q=0}^{\infty}\frac{(-1)^q\overline{\vartheta}^q_{(\nu,n)}}{q!(\overline{\Delta}_{(\nu)}+m_{cb}+q)}(1+w)^{\overline{\Delta}_{(\nu)}+m_{cb}}\frac{\Upsilon(\kappa+q+1,\Theta_{(\nu,ac)}x(w+1))}{\Theta^{\kappa+q+1}_{(\nu,ac)}}
	\end{align}
	and
	\begin{align}\label{psin2}
	&\overline{\varphi}_2(x|w)=K\!\!\sum_{\kappa=0}^{\varpi_{(\nu,ac)}}\frac{\zeta_{(\nu,ac)}(\kappa)}{\eta^{\kappa+1}_{a}}\sum_{n=0}^{K-1}\binom{K-1}{n}(-1)^{n}\alpha^{n+1}_{ac}\!\!\sum_{S_{(\nu,m)}\in\mathcal{S}}\!\frac{n!}{\prod_{m=0}^{\varpi_{(\nu,ac)}}s_m!}\prod_{m=0}^{\varpi_{(\nu,ac)}}\!(\mathcal{A}_{(\nu,m)})^{s_m}
	\\\nonumber
	&\times\sum_{v=0}^{n+1}\binom{n+1}{v}\frac{(-1)^v}{(\Gamma(m_{cb}))^v}\sum_{\overline{S}_p\in \overline{\mathcal{T}}}\frac{v!}{\prod_{p=0}^{\infty}\overline{s}_p!}\prod_{p=0}^{\infty}(\overline{\mathcal{B}}_p)^{\overline{s}_p}\frac{\Upsilon\left(\kappa+\overline{\Delta}_{(\nu)}+1,(n+1)\Theta_{(\nu,ac)}x(w+1)\right)}{(\left(n+1)\Theta_{(\nu,ac)}\right)^{\kappa+\overline{\Delta}_{(\nu)}+1}},
	\end{align}
	where $\overline{\vartheta}_{(\nu,n)} = n\Theta_{(\nu,ac)}+\frac{m_{cb}}{\Omega_{cb}\eta_c}$.
	\begin{IEEEproof}
		The proof follows Appendix \ref{appC} where we first invoke the terms $\overline{\varphi}_1(x|w)$ and $\overline{\varphi}_2(x|w)$ in place of ${\varphi}_1(x|w)$ and ${\varphi}_2(x|w)$, respectively, and then utilize (\ref{mu3}) to get (\ref{fac2}) after some straightforward mathematical steps.
	\end{IEEEproof}
	
	The OP of IoT network can now be derived as follows.
	\begin{theorem}\label{th1n}
		The lower bound on OP of the IoT network $\tilde{\mathcal{P}}^{\textmd{IoT}}_{\textmd{out}}(\mathcal{R}_{s})$ for Case 2 is given as
		\begin{align}\label{ajsl23n}
		\tilde{\mathcal{P}}^{\textmd{IoT}}_{\textmd{out}}(\mathcal{R}_{s})&=\left\{ \begin{array}{l}
		\overline{\Psi}_1(\mathcal{R}_{s}),\textmd{ if } \gamma_{s} < \frac{1}{\mu^{\prime}}, \\
		\overline{\Psi}_1(\mathcal{R}_{s})+\overline{\Psi}_2(\mathcal{R}_{s})+\overline{\Psi}_3(\mathcal{R}_{s}),\textmd{ if } \gamma_{s} \geq\frac{1}{\mu^{\prime}},
		\end{array}\right.
		\end{align}
		where $\overline{\Psi}_1(\mathcal{R}_{s})$ and $\overline{\Psi}_2(\mathcal{R}_{s})$ are given by 
		\begin{align}\label{si11n}
		\overline{\Psi}_1(\mathcal{R}_{s})&=\frac{1}{\Gamma(m_{cd})}\sum_{l=0}^{\infty}\frac{(-1)^l}{l! (m_{cd}+l)}\left(\frac{m_{cd}\gamma_s}{\Omega_{cd}\eta_c(1-\mu)}\right)^{m_{cd}+l}\widetilde{\sum_{(\nu,s)}}\,\frac{\Xi_{(\nu,s)}(M_s)}{\eta^{\Lambda}_s}\left(\frac{m_t}{\Omega_t\eta_t}\right)^{m_t M_t}\\\nonumber
		&\times\frac{\Phi(m_t M_t,\Lambda)}{\Gamma(m_t M_t)}\sum_{g=0}^{\infty}\frac{(\Lambda)_g (-\tilde{\Theta}_{(\nu,s)})^g}{(\tau(0))_g g!}\Gamma(\tau(g))U\left(\tau(g),\tau(g+m_{cd}+l+1); \frac{m_t}{\Omega_t \eta_t}\right)
		\end{align}
		and 
		\begin{align}\label{si22n}
		\overline{\Psi}_2(\mathcal{R}_{s})&=K\!\!\sum_{\kappa=0}^{\varpi_{(\nu,ac)}}\frac{\zeta_{(\nu,ac)}(\kappa)}{\eta^{\kappa+1}_{a}}\sum_{n=0}^{K-1}\binom{K-1}{n}(-1)^{n}\alpha^{n+1}_{ac}\sum_{S_{(\nu,m)}\in\mathcal{S}}\frac{n!}{\prod_{m=0}^{\varpi_{(\nu,ac)}}s_m!}\prod_{m=0}^{\varpi_{(\nu,ac)}}\\\nonumber
		&\times(\mathcal{A}_{(\nu,m)})^{s_m}\sum_{v=0}^{n+1}\binom{n+1}{v}\frac{(-1)^v}{(\Gamma(m_{cb}))^v}\sum_{\overline{S}_p\in \overline{\mathcal{T}}}\frac{v!}{\prod_{p=0}^{\infty}\overline{s}_p!}\prod_{p=0}^{\infty}(\overline{\mathcal{B}}_p)^{\overline{s}_p}\sum_{j=0}^{\infty}\frac{(-1)^j((n+1)\Theta_{(\nu,ac)})^j}{j!(\kappa+\overline{\Delta}_{(\nu)}+j+1)}\\\nonumber
		&\times\tilde{\gamma}^{\kappa+\overline{\Delta}_{(\nu)}+j+1}_s\widetilde{\sum_{(\nu,s)}}\,\frac{\Xi_{(\nu,s)}(M_s)}{\eta^{\Lambda}_s}\left(\frac{m_t}{\Omega_t\eta_t}\right)^{m_t M_t}\frac{\Phi(m_t M_t,\Lambda)}{\Gamma(m_t M_t)}\sum_{g=0}^{\infty}\frac{(\Lambda)_g (-\tilde{\Theta}_{(\nu,s)})^g}{(\tau(0))_g g!}\frac{\Gamma(\tau(g))}{\Gamma(m_{cd})}\\\nonumber
		&\times\left[\Gamma(m_{cd})U\left(\tau(g),\tau(g+\kappa+\overline{\Delta}_{(\nu)}+j+2); \frac{m_t}{\Omega_t \eta_t}\right)\right.-\sum_{l=0}^{\infty}\frac{(-1)^l}{l!(m_{cd}+l)}\\\nonumber
		&\times\left.\left(\frac{m_{cd}\gamma_s}{\Omega_{cd}\eta_c(1-\mu)}\right)^{m_{cd}+l}U\left(\tau(g),\tau(g+\kappa+\overline{\Delta}_{(\nu)}+m_{cd}+l+j+2); \frac{m_t}{\Omega_t \eta_t}\right)\right],
		\end{align}
		respectively. Also, $\overline{\Psi}_3(\mathcal{R}_{s})$ is given by  
		\begin{align}\label{si33n}
	\overline{\Psi}_3(\mathcal{R}_{s})	&=\frac{K}{\Gamma(m_{cb})}\left(\frac{m_{cb}}{\Omega_{cb}\eta_c}\right)^{m_{cb}}\sum_{n=0}^{K-1}\binom{K-1}{n}(-1)^{n}\alpha^{n+1}_{ac}\sum_{S_{(\nu,m)}\in\mathcal{S}}\frac{n!}{\prod_{m=0}^{\varpi_{(\nu,ac)}}s_m!}\\\nonumber
		&\times\prod_{m=0}^{\varpi_{(\nu,ac)}}(\mathcal{A}_{(\nu,m)})^{s_m}\sum_{v=0}^{n}\binom{n}{v}\frac{(-1)^v}{(\Gamma(m_{cb}))^v}\sum_{\overline{S}_p\in \overline{\mathcal{T}}}\frac{v!}{\prod_{p=0}^{\infty}\overline{s}_p!}\prod_{p=0}^{\infty}(\overline{\mathcal{B}}_p)^{\overline{s}_p}\sum_{\kappa=0}^{\varpi_{(\nu,ac)}}\frac{\zeta_{(\nu,ac)}(\kappa)}{\eta^{\kappa+1}_{a}}\\\nonumber
		&\times\sum_{q=0}^{\infty}\frac{(-1)^q\overline{\vartheta}^q_{(\nu,n)}}{q!(\overline{\Delta}_{(\nu)}+m_{cb}+q)}\frac{\Gamma(\kappa+q+1)}{{\Theta}^{\kappa+q+1}_{(\nu,ac)}}\widetilde{\sum_{(\nu,s)}}\,\frac{\Xi_{(\nu,s)}(M_s)}{\eta^{\Lambda}_s}\left(\frac{m_t}{\Omega_t\eta_t}\right)^{m_t M_t}\frac{\Phi(m_t M_t,\Lambda)}{\Gamma(m_t M_t)}\\\nonumber
		&\times\sum_{g=0}^{\infty}\frac{(\Lambda)_g (-\tilde{\Theta}_{(\nu,s)})^g}{(\tau(0))_g g!}\frac{\Gamma(\tau(g))}{\Gamma(m_{cd})}(\overline{\Psi}_4(\mathcal{R}_{s})-\overline{\Psi}_5(\mathcal{R}_{s})),
		\end{align}
		where $\overline{\Psi}_4(\mathcal{R}_{s})$ and $\overline{\Psi}_5(\mathcal{R}_{s})$ are given by 
					\begin{align}\label{si44n}
		\overline{\Psi}_4(\mathcal{R}_{s})&=\Gamma(m_{cd})\left[U\left(\tau(g),\tau(g+\overline{\Delta}_{(\nu)}+m_{cb}+1);\frac{m_t}{\Omega_t\eta_t}\right)\right.\\\nonumber
		&\left.-\sum_{u=0}^{\kappa+g}\frac{(\Theta_{(\nu,ac)}\tilde{\gamma}_s)^u}{u!}\textmd{e}^{-\Theta_{(\nu,ac)}\tilde{\gamma}_s}U\left(\tau(g),\tau(g+\overline{\Delta}_{(\nu)}+m_{cb}+u+1);\Theta_{(\nu,ac)}\tilde{\gamma}_s+\frac{m_t}{\Omega_t\eta_t}\right)\right]
		\end{align}
		and 
			\begin{align}\label{si55n}
		\overline{\Psi}_5(\mathcal{R}_{s})&=\sum_{l=0}^{\infty}\frac{(-1)^l}{l!(m_{cd}+l)}\left(\frac{m_{cd}\gamma_s}{\Omega_{cd}\eta_c(1-\mu)}\right)^{m_{cd}+l}\\\nonumber
		&\times\left[U\left(\tau(g),\tau(g+\overline{\Delta}_{(\nu)}+m_{cb}+m_{cd}+l+1);\frac{m_t}{\Omega_t\eta_t}\right)\right.-\sum_{u=0}^{\kappa+g}\frac{(\Theta_{(\nu,ac)}\tilde{\gamma}_s)^u}{u!}\\\nonumber
		&\times\left.\textmd{e}^{-\Theta_{(\nu,ac)}\tilde{\gamma}_s}U\left(\tau(g),\tau(g+\overline{\Delta}_{(\nu)}+m_{cb}+m_{cd}+l+u+1);\Theta_{(\nu,ac)}\tilde{\gamma}_s+\frac{m_t}{\Omega_t\eta_t}\right)\right],
		\end{align}
		respectively.
		\begin{IEEEproof}
			By first making use of Taylor series representation \cite[eq. 8.354.1]{grad} for function $\Upsilon(\cdot,\cdot)$ in (\ref{fsac}) and (\ref{psin2}) along with its finite series representation \cite[eq 8.352.6]{grad} in (\ref{psin1}) and then, evaluating (\ref{cm0sec1}) results in (\ref{ajsl23n}), where
			\begin{align}
			\overline{\Psi}_1(\mathcal{R}_{s})=\mathbb{E}(F_{\mu\hat{\Lambda}_{c_{k^{\ast}}d_{k^{\ast}}}}(\mu^{\prime}\gamma_{s}|w)),
			\end{align}
			\begin{align}
			\overline{\Psi}_2(\mathcal{R}_{s})=\mathbb{E}(\overline{\varphi}_2({\mu^{\prime}\gamma_{s}}-1|w)\overline{F}_{\mu\hat{\Lambda}_{c_{k^{\ast}}d_{k^{\ast}}}}(\mu^{\prime}\gamma_{s}|w)),
			\end{align}
			and
			\begin{align}
			\overline{\Psi}_3(\mathcal{R}_{s})=\mathbb{E}(\overline{\varphi}_1({\mu^{\prime}\gamma_{s}}-1|w)\overline{F}_{\mu\hat{\Lambda}_{c_{k^{\ast}}d_{k^{\ast}}}}(\mu^{\prime}\gamma_{s}|w)).
			\end{align}
			We can compute now $\overline{\Psi}_1(\mathcal{R}_{s})$, $\overline{\Psi}_2(\mathcal{R}_{s})$, and $\overline{\Psi}(\mathcal{R}_{s})$ similar to  $\mathcal{I}_2$ in Appendix \ref{appC}.
		\end{IEEEproof}
	\end{theorem}
		
		\subsubsection{Asymptotic OP} Next, we examine the asymptotic OP of IoT network for condition $(a)$.
		\begin{corollary}\label{cor2n}
			The asymptotic OP for secondary IoT network for Case $2$ under $\eta_a=\eta_c=\eta$  is given by 
			\begin{align}\label{asyy2n}
			\tilde{\mathcal{P}}^{\textmd{IoT}}_{\textmd{out},\infty}(\mathcal{R}_{s})&\simeq\left\{ \begin{array}{l}
			\frac{1}{\Gamma(m_{cd}+1)}\left(\frac{m_{cd}\gamma_{s}}{\Omega_{cd}\eta(1-\mu)}\right)^{m_{cd}} 
			\overline{\psi}_1({m}_{cd}), \textmd{ if } \gamma_{s} < \frac{1}{\mu^{\prime}},
			\\\frac{1}{\Gamma(m_{cd}+1)}\left(\frac{m_{cd}\gamma_{s}}{\Omega_{cd}\eta(1-\mu)}\right)^{m_{cd}} \overline{\psi}_1({m}_{cd})+\left(\frac{\alpha_{ac}\tilde{\gamma}_s}{\eta}\right)^K\psi_1(K), \textmd{ if } \gamma_{s} \geq \frac{1}{\mu^{\prime}} \textmd{ and } m_{cb} >1,
			\\\frac{1}{\Gamma(m_{cd}+1)}\left(\frac{m_{cd}\gamma_{s}}{\Omega_{cd}\eta(1-\mu)}\right)^{m_{cd}}
			\overline{\psi}_1({m}_{cd})
			+\frac{K}{m_{cd}(K-1)+1}\frac{\alpha_{ac}}{\eta}\tilde{\gamma}^{m_{cb}(K-1)+1}_s
			\\\times\left(\frac{1}{\Gamma(m_{cb}+1)}\left(\frac{m_{cb}}{\Omega_{cb}\eta}\right)^{m_{cb}}\right)^{K-1}\overline{\psi}_1({m}_{cb}(K-1)+1), \textmd{ if } \gamma_{s} \geq \frac{1}{\mu^{\prime}} \textmd{ and } m_{cb} <1,
			\end{array}\right.
			\end{align}
			where the function $\psi_1(x)$ and $\overline{\psi}_1(x)$ are the same as defined previously.
			\begin{IEEEproof}
				By following the proof of Corollary \ref{cor2}, one can obtain the cdf $F_{\hat{\Lambda}_{ac_{k^{\ast}}}}(x|w)$ for Case 2 at high SNR as
				\begin{align}\label{stgnh}
				F_{\hat{\Lambda}_{ac_{k^{\ast}}}}(x|w)\!&\simeq\!\left\{ \begin{array}{l}
				\left(\frac{\alpha_{ac} x }{\eta_c}\right)^K (w+1)^K, \textmd{ if } m_{cb} > 1, \\
				\frac{K}{m_{cb}(K-1)+1}\frac{\alpha_{ac}}{\eta}\left(\frac{1}{\Gamma(m_{cb}+1)}\right)^{K-1}  \left(\frac{m_{cb}}{\Omega_{cb}\eta}\right)^{m_{cb}(K-1)}x^{m_{cb}(K-1)+1} \\
				\times(w\!+\!1)^{m_{cb}(K-1)+1}, \textmd{ if } m_{cb} < 1,
				\end{array}\right.
				\end{align}
				Having the cdfs required in (\ref{aql23}), one can evaluate (\ref{cm0sec1}) by taking the final expectation as in Appendix \ref{appC} to get (\ref{asyy2n}).
			\end{IEEEproof}	 
		\end{corollary}
	
		The asymptotic OP of IoT network for condition $(b)$ is given below.
		\begin{corollary}\label{cor02n}
			The asymptotic OP of IoT network for Case 2 under $\eta_s=\eta_t=\lambda\eta$ is given by 
				\begin{align}\label{asyy02n}
			\tilde{\mathcal{P}}^{\textmd{IoT}}_{\textmd{out},\infty}(\mathcal{R}_{s})&\simeq\left\{ \begin{array}{l}
			\frac{1}{\Gamma(m_{cd}+1)}\big(\frac{m_{cd}\gamma_{s}}{\Omega_{cd}(1-\mu)}\big)^{m_{cd}} 
			\overline{\psi}_2({m}_{cd}), \textmd{ if } \gamma_{s} < \frac{1}{\mu^{\prime}},
			\\\frac{1}{\Gamma(m_{cd}+1)}\big(\frac{m_{cd}\gamma_{s}}{\Omega_{cd}(1-\mu)}\big)^{m_{cd}} \overline{\psi}_2({m}_{cd})+(\alpha_{ac}\tilde{\gamma}_s)^K\psi_2(K), \textmd{ if } \gamma_{s} \geq \frac{1}{\mu^{\prime}} \textmd{ and } m_{cb} >1,
			\\\frac{1}{\Gamma(m_{cd}+1)}\big(\frac{m_{cd}\gamma_{s}}{\Omega_{cd}(1-\mu)}\big)^{m_{cd}}
			\overline{\psi}_2({m}_{cd})
			+\frac{K}{m_{cd}(K-1)+1}\alpha_{ac}\tilde{\gamma}^{m_{cb}(K-1)+1}_s
			\\\times\left(\frac{1}{\Gamma(m_{cb}+1)}\left(\frac{m_{cb}}{\Omega_{cb}}\right)^{m_{cb}}\right)^{K-1}\overline{\psi}_2({m}_{cb}(K-1)+1), \textmd{ if } \gamma_{s} \geq \frac{1}{\mu^{\prime}} \textmd{ and } m_{cb} <1,
			\end{array}\right.
			\end{align}
			
			where the function $\psi_2(x)$ and $\overline{\psi}_2(x)$ are the same as defined in Corollary \ref{corr2} and \ref{cor1n1}, respectively.
			\begin{IEEEproof}
				Taking $\eta_s=\eta_t=\lambda\eta$ and following the proof of Corollary \ref{cor1n1}, we get (\ref{asyy02n}).
			\end{IEEEproof}
		\end{corollary}

\textit{Remark 3:} For Case 1 (i.e., INT/NINT SR and INT Nakagami-m fading), the diversity order for $\gamma_s<\frac{1}{\mu^{\prime}}$ and $\gamma_s\geq\frac{1}{\mu^{\prime}}$ is $m_{cd}$ and $\min(K,m_{cd})$ as seen from the asymptotic OP expression (\ref{asyy2}) under condition $(a)$ for IoT network. Similar to Case 1, the diversity order for Case 2 under condition $(a)$ and $\gamma_s<\frac{1}{\mu^{\prime}}$ is $m_{cd}$. However, for $\gamma_s\geq\frac{1}{\mu^{\prime}}$ under condition $(b)$, the diversity order is $\min(K,m_{cd})$ and $\min(m_{cb}(K-1)+1,m_{cd})$ for $m_{cb}>1$ and $m_{cb}<1$, respectively, according to the OP expression in (\ref{asyy2n}). Similar to satellite network, the diversity order reduces to zero for IoT network as seen from (\ref{asy2}) and (\ref{asyy02n}) under condition $(b)$.

	\section{Adaptive Power-Splitting Factor}\label{psf}
	In this section, we devise the scheme for finding the appropriate value of power-splitting factor $\mu$ for effective spectrum sharing. Recalling the necessary condition $\gamma_{p} < \mu^{\prime}$ in Theorem \ref{tth2}, the feasible dynamic range of $\mu$ can be formulated as $\frac{\gamma_{\textmd{p}}}{1+\gamma_{\textmd{p}}} \leq \mu \leq 1$. Further, to obtain $\mu$, a QoS constraint must be imposed to protect the satellite network from IoT transmissions. Thus, we choose the value of $\mu$ such that the OP of the satellite network $\tilde{\mathcal{P}}^{\textmd{sat}}_{\textmd{out}}(\mathcal{R}_{p})$ is guaranteed below a predetermined QoS level $\epsilon$, i.e., $\tilde{\mathcal{P}}^{\textmd{sat}}_{\textmd{out}}(\mathcal{R}_{p})\leq \epsilon$. Note that if this QoS constraint is taken at equality, the resulting value of $\mu$ minimizes the OP of IoT network (i.e., $\tilde{\mathcal{P}}^{\textmd{IoT}}_{\textmd{out}}(\mathcal{R}_{s})$). Although the closed-form solution of $\mu$ under above constraints is infeasible, it can be determined via numerical search method. Moreover, we consider the case of assigning an arbitrary fixed value of $\mu$ within its dynamic range for comparison. 
	
	\section{Numerical and Simulation Results}\label{num}
In this section, we present numerical results for considered OSTN. Here, the simulations are conducted for $10^6$ independent channel realizations. We set $\mathcal{R}_{p}=\mathcal{R}_{s}=0.5$ bps/Hz such that $\gamma_{p}=\gamma_{s}=1$ (unless stated otherwise). We consider $M_s=M_t=2$ and $M_s=M_t=1$ for obtaining the results under fixed and adaptive values of $\mu$, respectively. The fixed value of $\mu$ is set as $0.75$. Further, we set $\Omega_{cb}=\Omega_{cd}=1$ and $\eta_{a}=\eta_{c}=\eta$ as SNR. We consider the SR fading parameters of satellite link $A-C_{k}$ for INT and NINT as ($m_{ac},\flat_{ac},\Omega_{ac}=5,0.251,0.279$) (for light shadowing) and ($m_{ac},\flat_{ac},\Omega_{ac}=1.95,0.063,0.0005$) (for heavy shadowing), respectively. The respective INT and NINT SR fading parameters of interfering ETSs are  ($m_{s},\flat_{s},\Omega_{s}=1,0.063,0.0005$) and ($m_{s},\flat_{s},\Omega_{s}=0.95,0.063,0.0005$) for heavy shadowing. We further set $\Omega_t=0.1$ for interfering TSs. The powers of interfering ETSs and TSs are set as $\eta_s=\eta_t=1$ dB. The Nakagami-\emph{m} fading parameters corresponding to interfering TSs are set $m_t=2$ and $m_t=1.77$ for Cases 1 and 2, respectively. Further, for condition $(a)$, we set $\eta_s=\eta_t=1$ dB; and for condition $(b)$, we set $\eta_s=\eta_t=\lambda\eta$, with $\lambda=-20$ dB. Moreover, we consider the relevant Scenarios 1, 2, 3 and 4 as described in Table \ref{tab}.
	
	\subsection{OP of Satellite Network: Fixed $\mu$, Condition (a)}
\begin{figure}[!t]
		\begin{minipage}{.5\textwidth}
		\centering
		\includegraphics[width=3.2in]{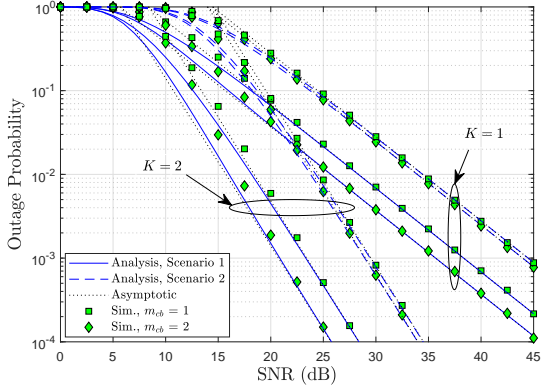}
		\caption{OP of satellite network versus SNR for Case 1.}
		\label{fig1}
	\end{minipage}%
\begin{minipage}{.5\textwidth}
		\centering
		\includegraphics[width=3.2in]{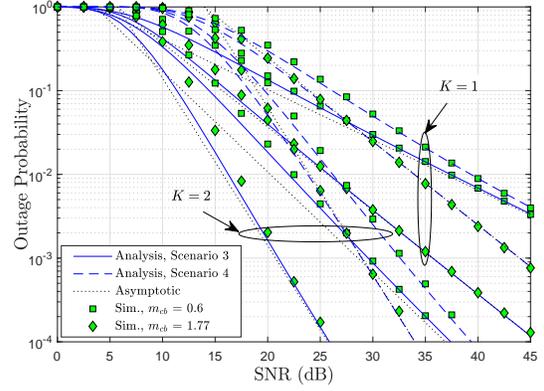}
		\caption{OP of satellite network versus SNR for Case 2.}
		\label{fig2}
	\end{minipage}
	\end{figure}

	In Figs.~\ref{fig1} and \ref{fig2}, we plot the OP curves of satellite network for Cases 1 and 2, respectively. Further, in Fig.~\ref{fig1}, the results correspond to the Scenarios 1 and 2. Whereas, in Fig.~\ref{fig2}, the curves are obtained for Scenarios 3 and 4. We can clearly observe that for given values of parameters $\eta_s$, $\eta_t$ and $\mu$, our analytical lower bound OP curves are in close proximity to the exact simulation results. Also, the effectiveness of our theoretical analysis can be observed for the Cases $1$ and $2$ in Figs.~\ref{fig1} and \ref{fig2}, respectively. Furthermore, the asymptotic curves at high SNR are well-aligned with the analytical and simulation results. Specifically, in Fig.~\ref{fig1}, if $m_{cb}$ changes from $1$ to $2$ keeping the $K$ fixed, the slope of OP curves remain unchanged. However, when $K$ changes from $1$ to $2$ irrespective of the value of parameter $m_{cb}$, the slope of the OP curves increases. Thus, a diversity order of $K$ is achievable for satellite network under Case 1. Unlike Scenario 1, in Scenario 2, the acceptable tightness can be achieved by truncating the infinite series to $10$ terms only. Thus, our proposed solution for Scenario 2 is quite efficient. Different from Fig.~\ref{fig1}, in Fig.~\ref{fig2}, the achievable diversity order of $K$ and $K m_{cb}$ is attested for $m_{cb}<1$ and $m_{cb}>1$, respectively. It is apparent from the slope of curves with different values of $m_{cb}$ for fixed $K$, i.e., $1$ or $2$. More importantly, it is observed that as the SNR increases, the OP becomes independent of SR fading parameters when $m_{cb}<1$. Here, both Scenarios 3 and 4 have infinite series solutions which converge effectively with 20 terms.	 

  \subsection{OP of IoT Network: Fixed $\mu$, Condition (a)}
\begin{figure}[!t]
		\begin{minipage}{.5\textwidth}
		\centering
		\includegraphics[width=3.2in]{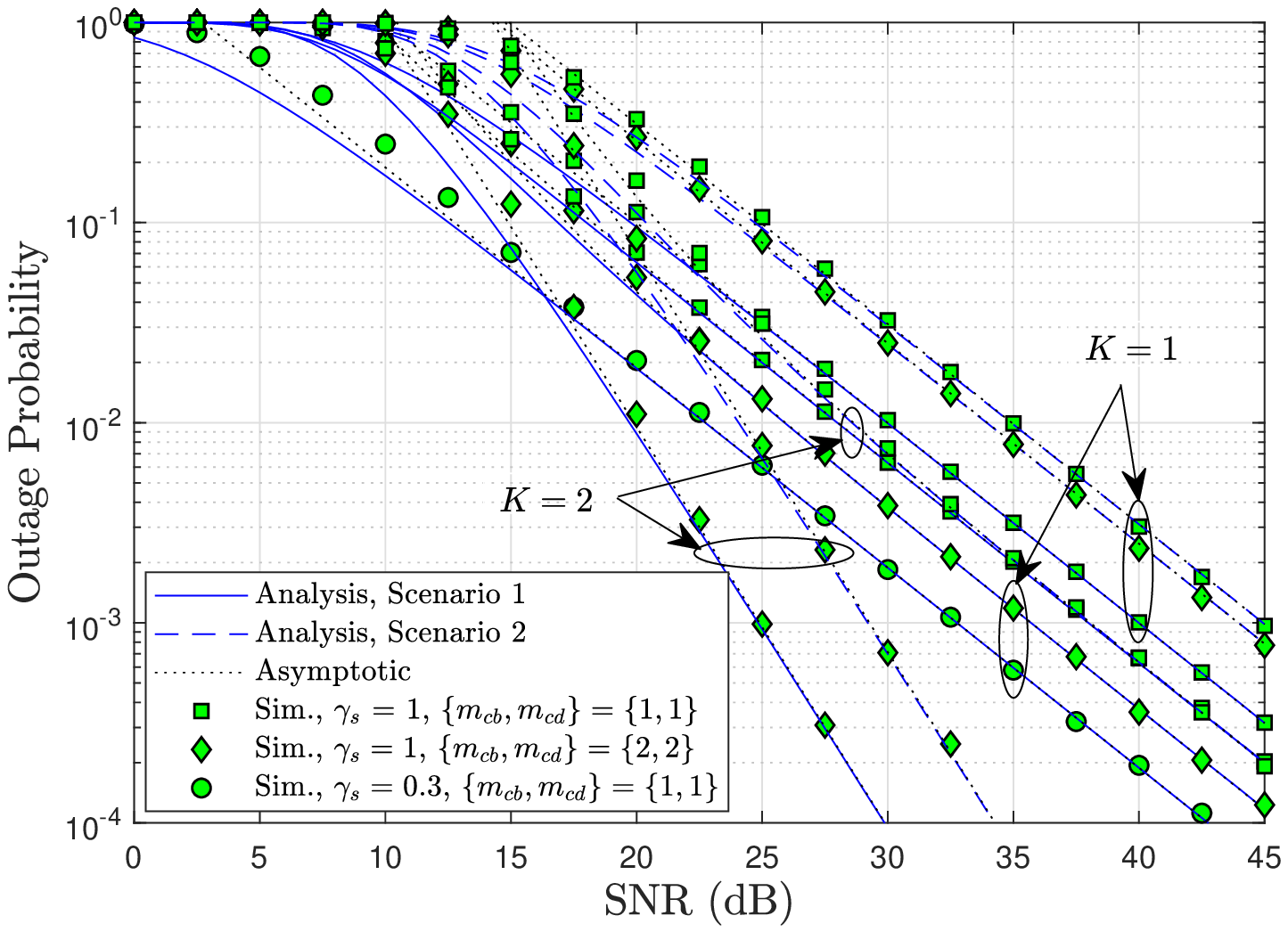}
		\caption{OP of IoT network versus SNR for Case 1.}
		\label{fig3}
	\end{minipage}%
\begin{minipage}{.5\textwidth}
		\centering
		\includegraphics[width=3.2in]{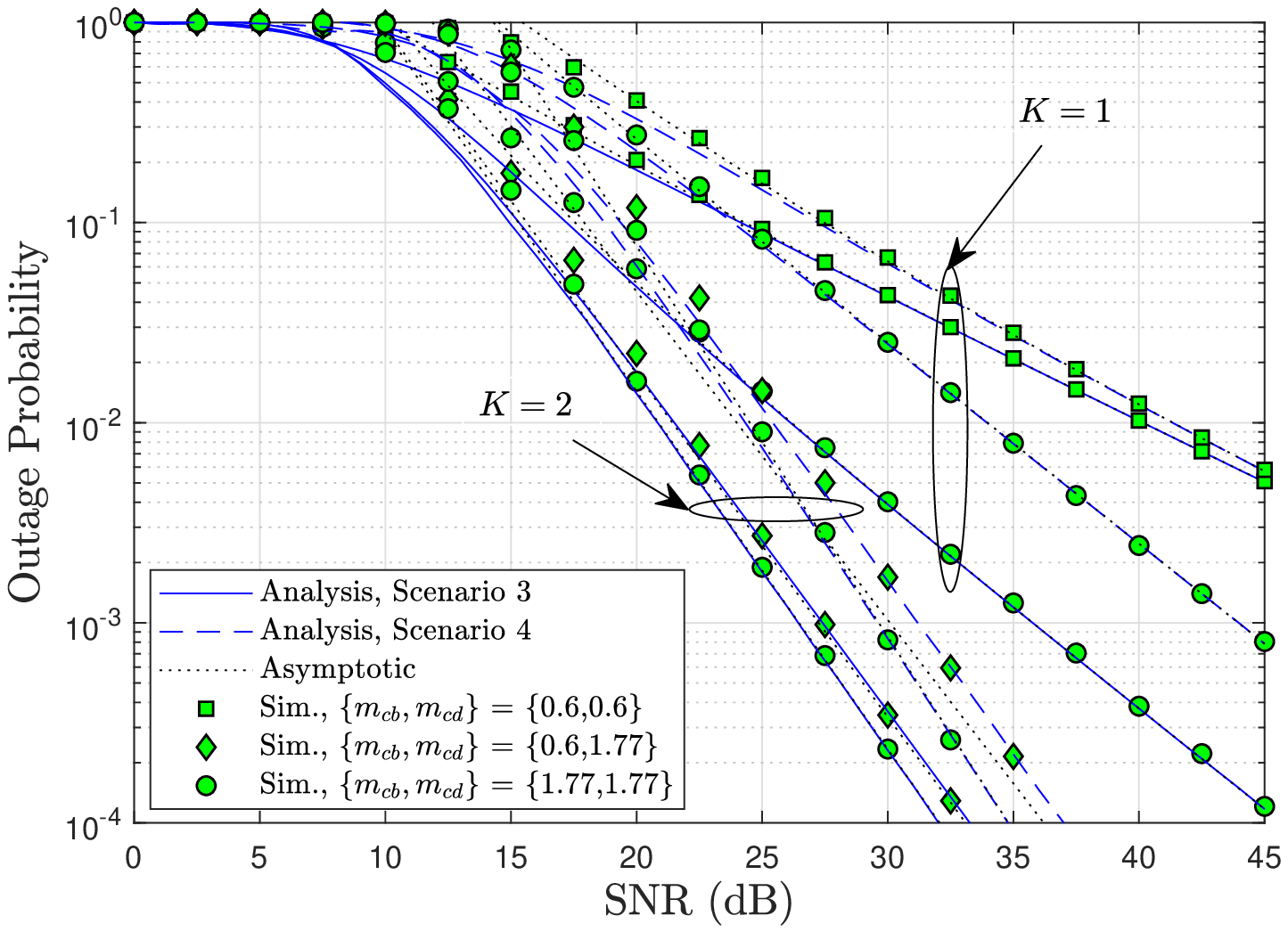}
		\caption{OP of IoT network versus SNR for Case 2.}
		\label{fig4}
	\end{minipage}%
	\end{figure}

	In Figs.~\ref{fig3} and \ref{fig4}, we plot the OP curves of IoT network for Cases 1 and 2, respectively. Likewise, in Fig.~\ref{fig3} and Fig.~\ref{fig4}, we plot the curves corresponding to Scenarios 1 and 2 and Scenarios 3 and 4, respectively. Clearly, for given values of $\eta_s$, $\eta_t$ and $\mu$, we can observe that the lower bound analytical OP curves are well-aligned with the simulation results. It is applicable for both Cases 1 and 2 in respective Figs.~\ref{fig3} and \ref{fig4}. Specifically, in Fig.~\ref{fig3}, we set $\gamma_s=1$ and $0.3$ corresponding to conditions $\gamma_s\geq\frac{1}{\mu^\prime}$ and $\gamma_s<\frac{1}{\mu^\prime}$ as per Theorem \ref{th1}. Here, when $K$ changes from $1$ and $2$ with $\{m_{cb}, m_{cd}\}=\{1,1\}$ and $\{2,2\}$, the slope of various OP curves reveal the diversity order of $\min(K,m_{cd})$ for IoT network under Case 1. Herein, in Scenario 2, it takes 20 terms for series convergence to achieve acceptable tightness of results. However, in Fig.~\ref{fig4}, two different diversity orders are observed for IoT network, i.e., $\min(K,m_{cd})$ for $m_{cb}<1$ and $\min(m_{cb}(K-1)+1,m_{cd})$ for $m_{cb}>1$ under Case 2. Note that the achievable diversity orders depend upon the combination of multiple parameters. For instance, when $(K=2,m_{cb}=1.77,m_{cd}=1.77)$, the achievable diversity order is $1.77$ and when $(K=2,m_{cb}=0.6,m_{cd}=1.77)$, the diversity order changes to $1.6$. Another important observation is that as SNR increases, the OP of IoT network becomes independent of SR fading parameters for $m_{cb}<1$. Further, for the convergence of infinite series under Scenario 3 and 4, there is a requirement of $200$ and $30$ terms at SNRs $0$dB and $10$ dB, respectively. Thus, at high SNR, the number of terms required for series convergence is significantly lesser. Moreover, the performance of IoT network has dependence on the fading of $C_k-B$ link necessary for primary satellite communications.

\subsection{OP of Satellite Network: Fixed $\mu$, Condition (b)}
\begin{figure}[!t]
		\begin{minipage}{.5\textwidth}
		\centering
		\includegraphics[width=3.2in]{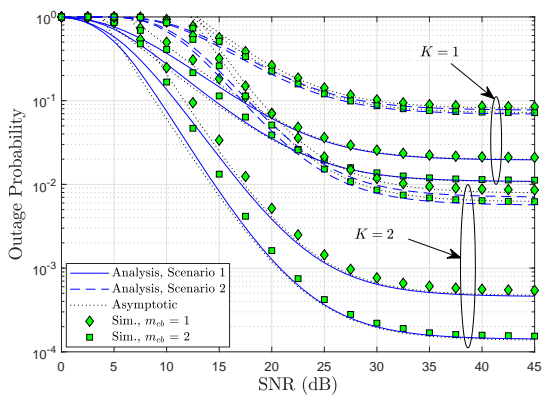}
		\caption{OP of satellite network versus SNR for Case 1.}
		\label{fig5}
		\end{minipage}%
\begin{minipage}{.5\textwidth}
		\centering
		\includegraphics[width=3.2in]{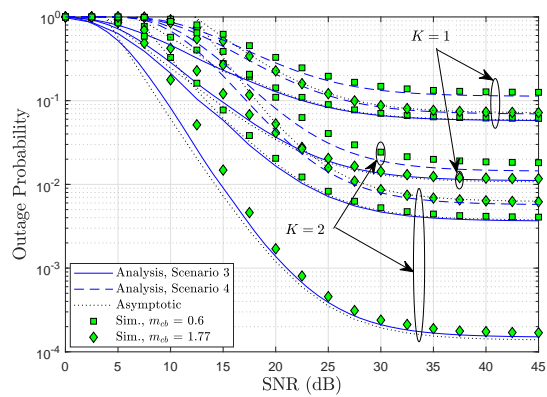}
		\caption{OP of satellite network versus SNR for Case 2.}
		\label{fig6}
	\end{minipage}%
	\end{figure}

		In Figs.~\ref{fig5} and \ref{fig6}, we plot the OP curves of the satellite network for Cases 1 and 2, respectively. Here, Figs.~\ref{fig5} and \ref{fig6}, the curves are plotted for Scenarios 1 and 2 and Scenarios 3 and 4, respectively. 
		We observe that unlike the fixed power interferers in Figs.~\ref{fig1} and \ref{fig2}, hereby the achievable diversity order of satellite network reduces to zero. This is apparent through zero-slope flat OP curves at high SNR irrespective of underlying system and channel parameters. It follows due to difficulty in maintaining the high SNR at $B$ when interferers increase their transmit power. However, in both these figures, we can see that the outage performance of satellite network can still be improved when $K$ increases from $1$ to $2$ and/or $m_{cb}$ from $1$ to $2$. This improvement in the outage performance results due to coding gain only. It is worth mentioning that a careful choice of $K$ for given $m_{cb}$ helps realizing the performance gain.     

	\subsection{OP of IoT Network: Fixed $\mu$, Condition (b)}
\begin{figure}[!t]
			\begin{minipage}{.5\textwidth}
		\centering
		\includegraphics[width=3.2in]{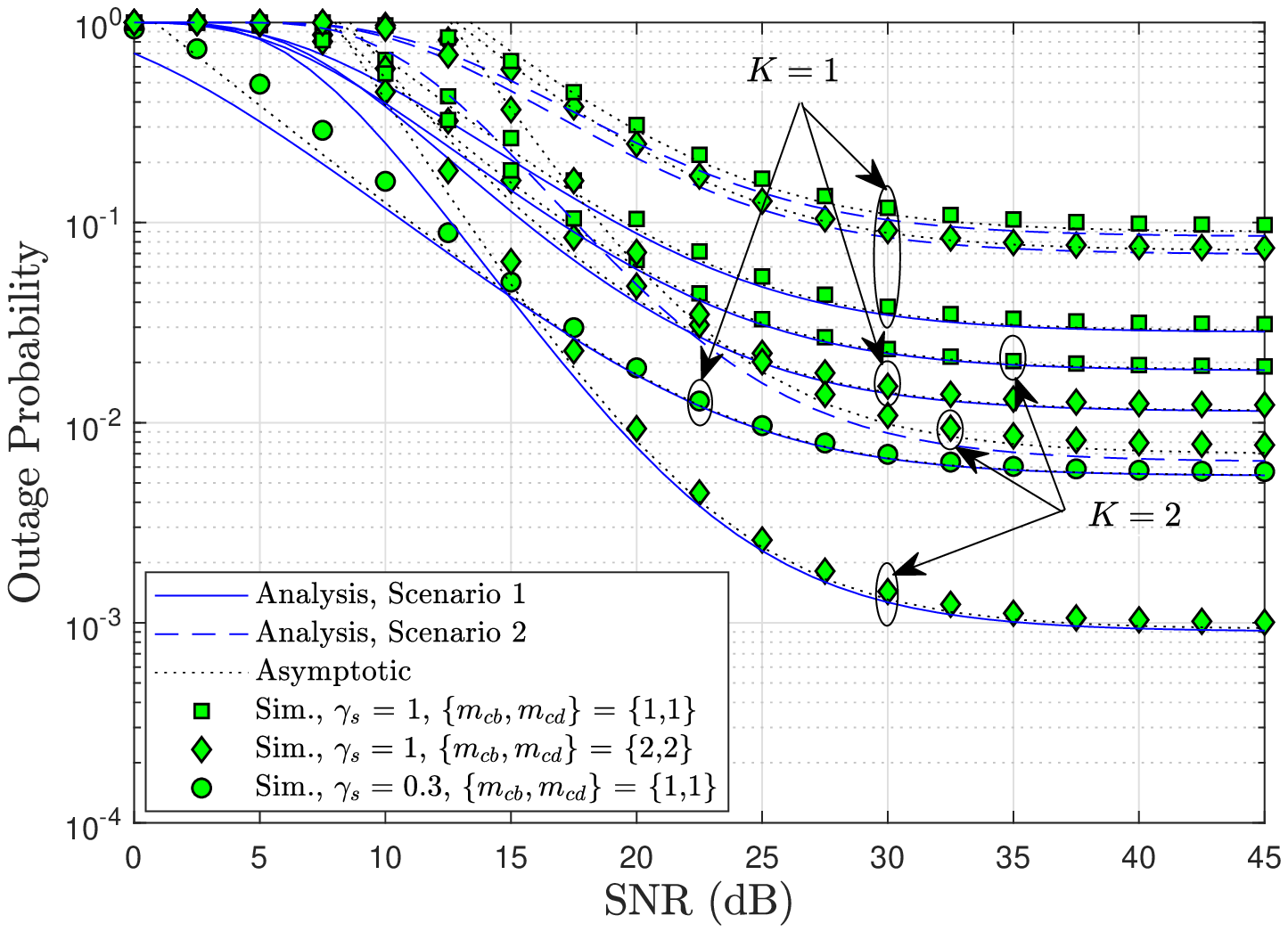}
		\caption{OP of IoT network versus SNR for Case 1.}
		\label{fig7}
		\end{minipage}%
\begin{minipage}{.5\textwidth}
		\centering
		\includegraphics[width=3.2in]{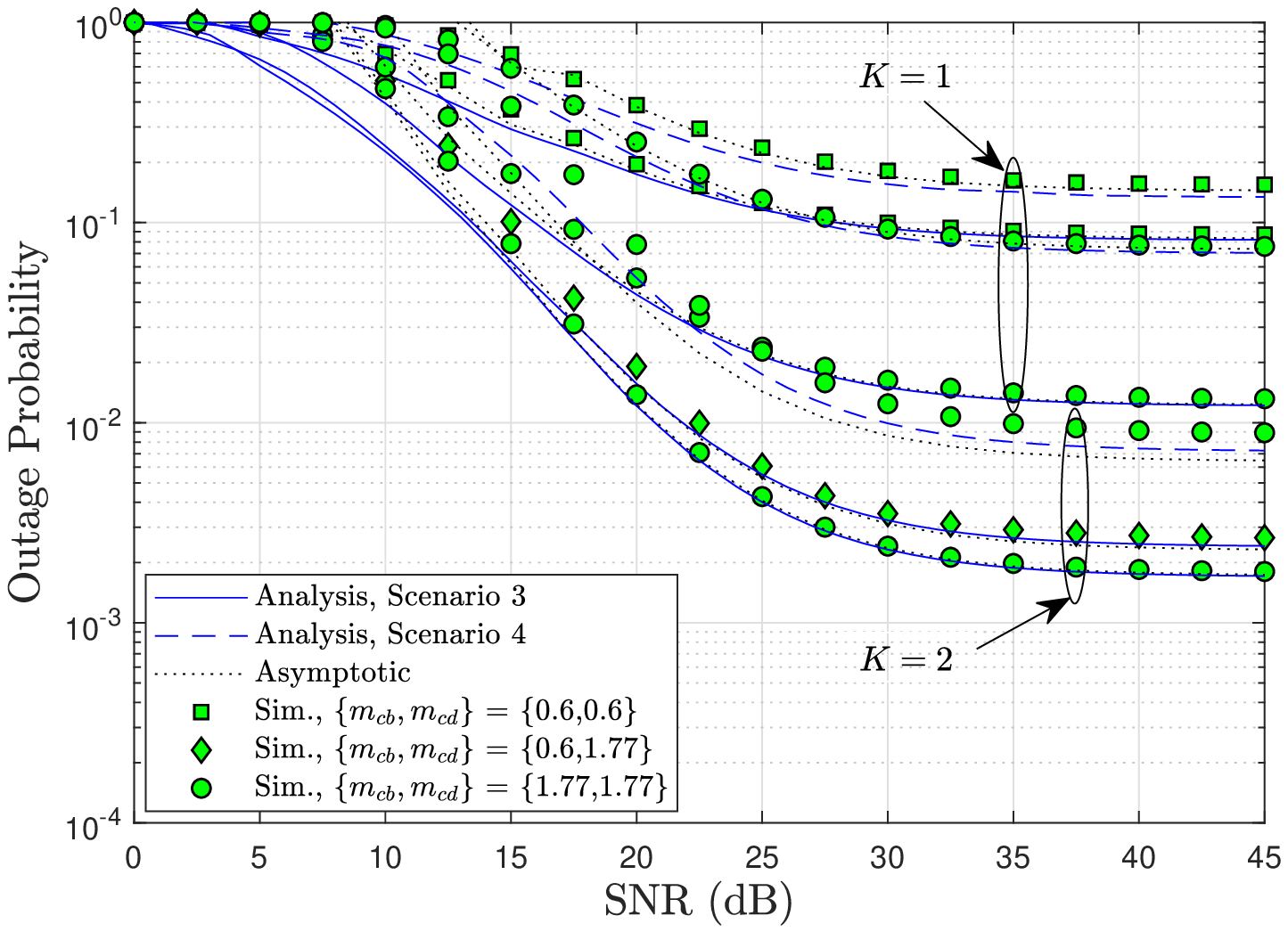}
		\caption{OP of IoT network versus SNR for Case 2.}
		\label{fig8}
	\end{minipage}%
	\end{figure}

	In Figs.~\ref{fig7} and \ref{fig8}, we plot the OP curves of the IoT network for Cases 1 and 2, respectively. Here, in Fig.~\ref{fig7}, the plots correspond to Scenarios 1 and 2 while in Fig.~\ref{fig8}, the plots correspond to Scenarios 3 and 4. 
	Apparently, as compared to the situations in Figs.~\ref{fig3} and \ref{fig4}, hereby, the achievable diversity order of IoT network reduces to zero. It can be seen through the flat OP curves in high SNR region irrespective of the choice of parameters $\{K,m_{cb}, m_{cd}\}$. However, similar to that for satellite network, in these figures, the outage performance of IoT network is shown to  improve when $K$ and/or $m_{cb}$ increases due to achievable coding gain. 
	
	\subsection{OP of IoT Network: Adaptive $\mu$, Condition (a)}
\begin{figure}[!t]
			\begin{minipage}{.5\textwidth}
		\centering
		\includegraphics[width=3.2in]{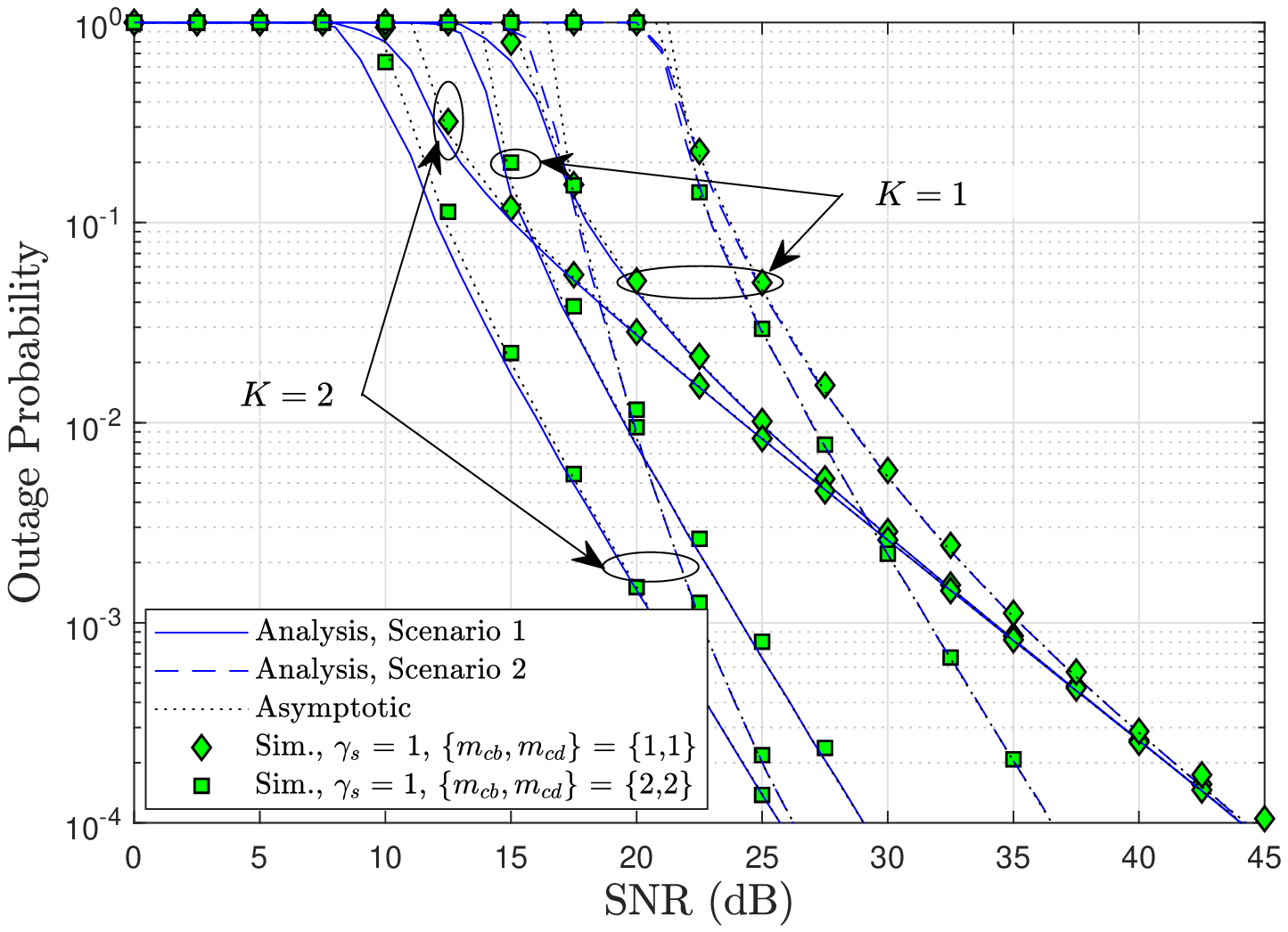}
		\caption{OP of IoT network versus SNR for Case 1.}
		\label{fig9}
		\end{minipage}%
\begin{minipage}{.5\textwidth}
		\centering
		\includegraphics[width=3.2in]{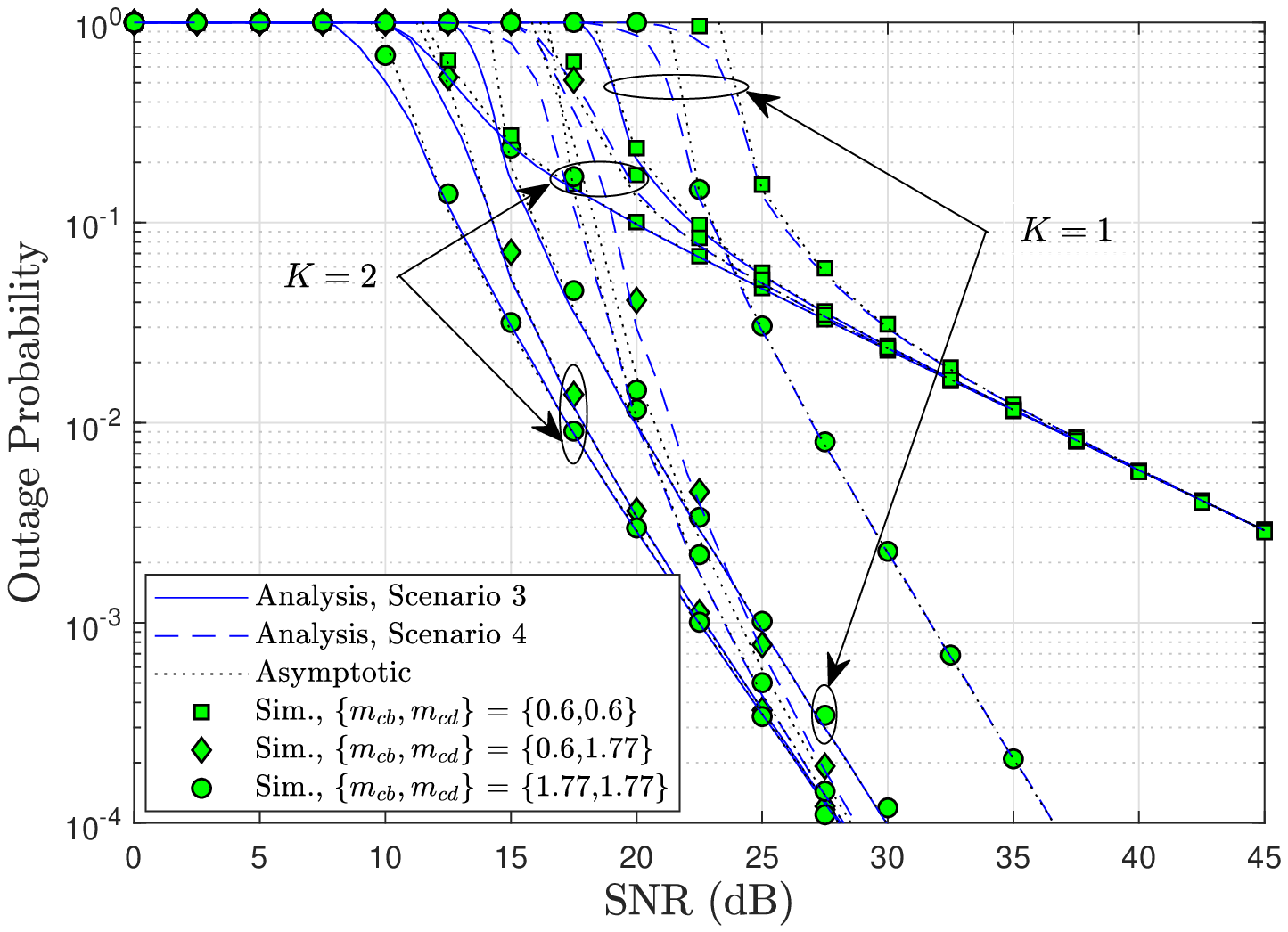}
		\caption{OP of IoT network versus SNR for Case 2.}
		\label{fig10}
	\end{minipage}%
	\end{figure}

	In Figs.~\ref{fig9} and \ref{fig10}, we plot the OP curves of the IoT network for Cases 1 and 2, respectively, under adaptive $\mu$ (please refer to Section \ref{psf}). For this, we set $M_s=M_t=1$ and outage threshold as $10\%$, i.e., $\epsilon=0.1$. Note that, hereby, the OP of the IoT network remains unity up to certain SNR level until the primary QoS constraint is not satisfied, i.e., $\tilde{\mathcal{P}}^{\textmd{sat}}_{\textmd{out}}(\mathcal{R}_{p})\leq \epsilon$. Up to the range of this SNR, both the satellite and IoT are jointly experiencing the signal outage since the required QoS constraint is not met. However, beyond this SNR level, the OP of IoT network improves remarkably with adaptive $\mu$. We further comment that this behaviour applies to Scenarios 1 and 2 in Fig.~\ref{fig9} as well as to Scenarios 3 and 4 in Fig.~\ref{fig10}. 
	Furthermore, although not shown explicitly, the value of $\mu$ was found to approach its minimum possible value, i.e., $0.5$, as SNR increases. As a result, IoT network gets a higher fraction of its available power for secondary communication. Consequently, the outage performance of IoT network improves by simultaneously protecting the QoS of satellite network.     
	
	\subsection{OP of IoT Network: Adaptive $\mu$, Condition (b)}
\begin{figure}[!t]
		\begin{minipage}{.5\textwidth}
		\centering
		\includegraphics[width=3.2in]{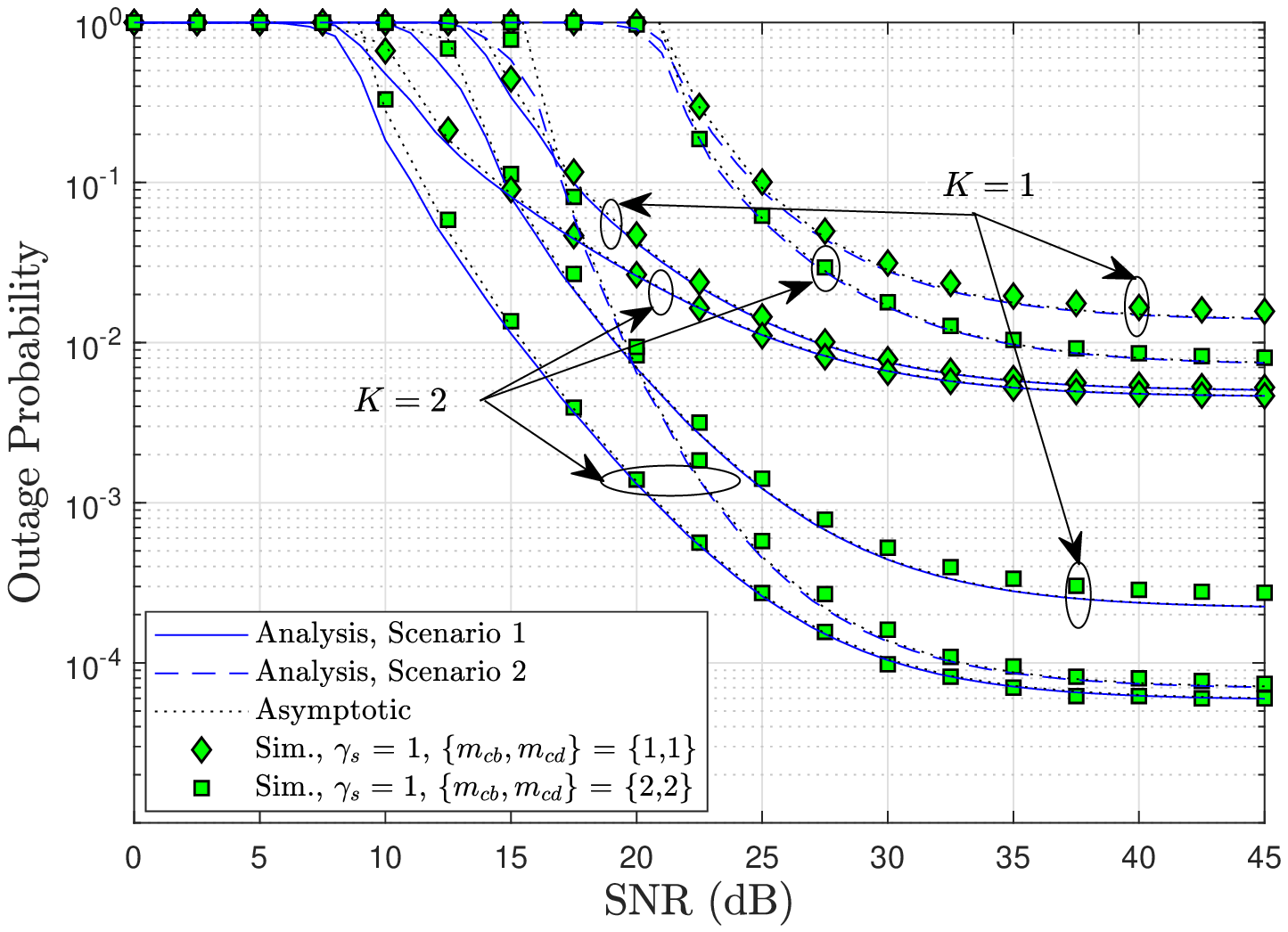}
		\caption{OP of IoT network versus SNR for Case 1.}
		\label{fig11}
	\end{minipage}%
\begin{minipage}{.5\textwidth}
		\centering
		\includegraphics[width=3.2in]{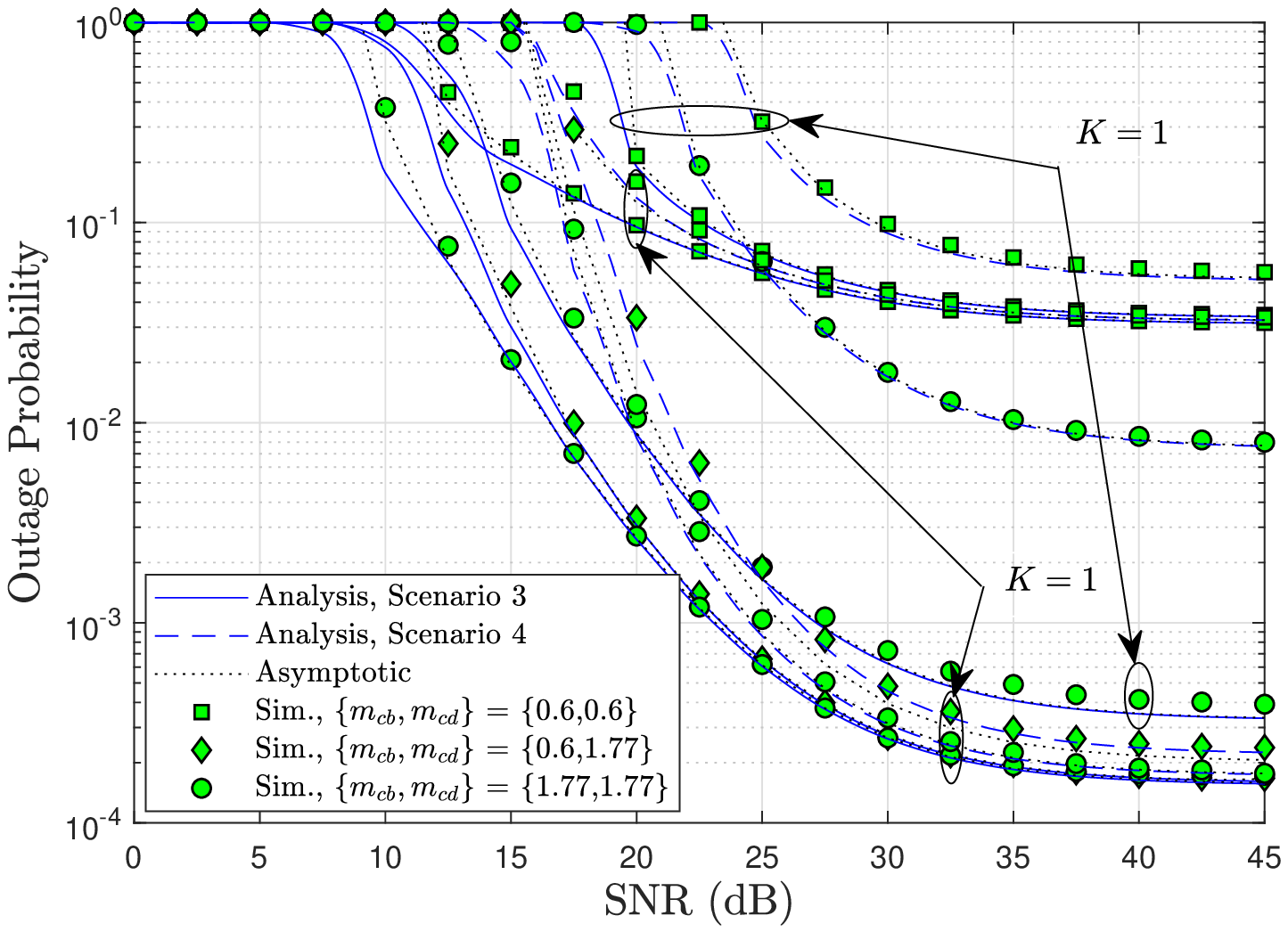}
		\caption{OP of IoT network versus SNR for Case 2.}
		\label{fig12}
	\end{minipage}%
	\end{figure}

	In Figs.~\ref{fig11} and \ref{fig12}, we plot the OP curves of the IoT network for Cases 1 and 2, respectively, under adaptive $\mu$. Here, we set $M_s=M_t=1$, and $\epsilon=0.1$. As described previously, herein, the satellite and IoT networks remain in outage until certain SNR level. It is due to violation of the QoS constraint imposed by the satellite network. Further, different from Figs.~\ref{fig9} and \ref{fig10}, the achievable diversity order of IoT network is zero. However, the coding gain can still be harnessed for performance gain under appropriate choice of parameters $\{K,m_{cb}, m_{cd}\}$.  
	While plotting the curves, we have observed the behaviour of adaptive values of $\mu$ as SNR increases. It has observed that $\mu$ may not always approach to its minimum value $0.5$ even at very high SNR. For instance, when $K=1,m=1$, $\mu$ converges to $0.523$ for Scenario 1 under Case 1. Likewise, when $K=1,m=0.6$, $\mu$ converges to $0.553$ for Scenario 3 under Case 2. This is due to the fact that OP takes on constant values (i.e., zero-slope flat OP curves) for entire high SNR regime under the condition $(b)$. 

	\section{Conclusion}\label{con}
In this paper, we have analyzed the OP of an OSTN where a selected secondary IoT network assists primary satellite communications in the presence of hybrid interference from ETSs and TSs. We presented a unified framework for INT/NINT Nakagami-\emph{m} parameter of SR fading related to main satellite and interfering ETSs links. In addition, we considered both INT/NINT Nakagami-\emph{m} fading scenario for main terrestrial and interfering TSs. We derived tight lower bound OP expressions for both satellite and IoT networks under two cases 1 and 2 which characterize all four scenarios described in Table \ref{tab}. We further derived asymptotic OP expressions for both these networks under the conditions $(a)$ and $(b)$ to find their achievable diversity orders. We have also formulated an adaptive scheme for power splitting-factor that improves the OP of IoT networks while guaranteeing certain QoS of satellite network. In general, we found that in the presence of hybrid interference, the achievable diversity orders of both the networks are different under the cases 1 and 2. The diversity orders of these networks under condition $(a)$ depend upon the  choice of INT/NINT combination of parameters for both SR and Nakagami-\emph{m} fading. However, under condition $(b)$, the achievable diversity order of both these networks reduced to zero irrespective of the INT/NINT SR and Nakagami-\emph{m} fading parameters. We found that even when the diversity orders of the satellite and IoT networks become zero, the coding gain can be harnessed for enhancing their system performance. Nevertheless, our generalized OP analysis of considered OSTN for general INT/NINT SR and Nakagami-\emph{m} parameters has laid  useful guidelines for futuristic deployments.  
	\appendices
	\section{Proof of Theorem \ref{tth2}}\label{appA}
	To derive (\ref{cmqr}), we first apply the bound $\frac{XY}{X+Y}\leq\min(X,Y)$ to re-express (\ref{snrs}) as  
	\begin{align}\label{hisn}
	{\Lambda}_{ac_{k}b}&\leq{\Lambda}^{\star}_{ac_{k}b}=\frac{\mu}{(1-\mu)+1/\min({\hat{\Lambda}_{ac_{k}},\hat{\Lambda}_{c_{k}b}})}.
	\end{align} 
	Eventually, we substitute this upper bound SINR ${\Lambda}^{\star}_{ac_{k}b}$ in (\ref{cmqr}) to get 
	\begin{align}\label{hsi}
	\tilde{\mathcal{P}}^{\textmd{sat}}_{\textmd{out}}(\mathcal{R}_{p})&=\mathbb{E}\{(1-\overline{F}_{\hat{\Lambda}_{ac_k}}(\tilde{\gamma}_p|w)\overline{F}_{\hat{\Lambda}_{c_kb}}(\tilde{\gamma}_p|w))^K\}
	\end{align}
	which after applying the binomial expansion results in
	\begin{align}\label{hrsi}
	\tilde{\mathcal{P}}^{\textmd{sat}}_{\textmd{out}}(\mathcal{R}_{p})&=
	\mathbb{E}\{\sum_{n=0}^{K} \binom{K}{n}(-1)^n[\overline{F}_{\hat{\Lambda}_{ac_k}}(\tilde{\gamma}_p|w)]^n[\overline{F}_{\hat{\Lambda}_{c_kb}}(\tilde{\gamma}_p|w)]^n\},
	\end{align}
	where $\overline{F}_{X}(\cdot|w)=1-F_{X}(\cdot|w)$. Further, based on the Multinomial expansion \cite{pkss}, we can express
	\begin{align}\label{mu1}
	&[\overline{F}_{\hat{\Lambda}_{ac_k}}(x|w)]^n = \alpha^n_{ac}\sum_{S_{(\nu,m)}\in\mathcal{S}}\frac{n!}{\prod_{m=0}^{\varpi_{(\nu,ac)}}s_m!}\prod_{m=0}^{\varpi_{(\nu,ac)}}(\mathcal{A}_{(\nu,m)})^{s_m}\left(x(w+1)\right)^{\Delta_{(\nu,ac)}}\textmd{e}^{-{\Theta}_{(\nu,ac)}nx(w+1)}
	\end{align}
	and
	\begin{align}\label{mu2}
	[\overline{F}_{\hat{\Lambda}_{c_kb}}(x|w)]^n = &\sum_{S_p\in \mathcal{T}_1}\frac{n!}{\prod_{p=0}^{m_{cb}-1}s_p!}\prod_{p=0}^{m_{cb}-1}(\mathcal{B}_p)^{s_p}\left(x(w+1)\right)^{\Delta_{cb}}\textmd{e}^{-\left(\frac{m_{cb}}{\Omega_{cb}\eta_{c}}\right)nx(w+1)}.
	\end{align}
	
	Next, on invoking (\ref{mu1}) and (\ref{mu2}) in (\ref{hrsi}), we can reach
	\begin{align}\label{62}
	&\tilde{\mathcal{P}}^{\textmd{sat}}_{\textmd{out}}(\mathcal{R}_{p})=\sum_{n=0}^{K}\binom{K}{n}(-1)^n\alpha^n_{ac}\sum_{S_{(\nu,m)}\in\mathcal{S}}\frac{n!}{\prod_{m=0}^{\varpi_{(\nu,ac)}}s_m!}\prod_{m=0}^{\varpi_{(\nu,ac)}}(\mathcal{A}_{(\nu,m)})^{s_m}\sum_{S_p\in \mathcal{T}_1}\frac{n!}{\prod_{p=0}^{m_{cb}-1}s_p!}\\\nonumber
	&\times\prod_{p=0}^{m_{cb}-1}(\mathcal{B}_p)^{s_p}\tilde{\gamma}^{\Delta_{(\nu)}}_p\textmd{e}^{-\tilde{\Theta}_{(\nu,ac)} n \tilde{\gamma}_{p}}\underbrace{\mathbb{E}\{(w+1)^{\Delta_{(\nu)}}\textmd{e}^{-\tilde{\Theta}_{(\nu,ac)} n \tilde{\gamma}_{p}w}\}}_{\mathcal{I}_1}.
	\end{align}
	Furthermore, the term $\mathcal{I}_1$ in (\ref{62}) can be re-expressed, by making use of pdf of $W_s$ from (\ref{wct}), as
	\begin{align}\label{63}
	\mathcal{I}_1&=\widetilde{\sum_{(\nu,s)}}\,\frac{\Xi_{(\nu,s)}(M_s)}{\eta^{\Lambda}_s}\left(\frac{m_t}{\Omega_t\eta_t}\right)^{m_t M_t}\frac{\Phi(m_t M_t,\Lambda)}{\Gamma(m_t M_t)}\int_{0}^{\infty}w^{\Lambda+m_tM_t-1}\left(w+1\right)^{\Delta_{(\nu)}}\\\nonumber
	&\times\textmd{e}^{-\left(\tilde{\Theta}_{(\nu,ac)}n\tilde{\gamma}_{p}+\frac{m_t}{\Omega_t \eta_t}\right)w}{}_1F_1(\Lambda;m_tM_t;-\tilde{\Theta}_{(\nu,s)}w)dw.
	\end{align}
	Finally, upon applying the binomial expansion for $(w+1)^{\Delta_{(\nu)}}$ and evaluating the resulting integral with the aid of \cite[eq. 7.621.4]{grad}, we can achieve (\ref{asl23}).
	
	\section{Proof of Corollary \ref{cor1}}\label{appB}
	First, we approximate (\ref{hsi}) in Appendix \ref{appA} by neglecting the higher-order term given by the product of two conditional cdfs (i.e, $F_{\hat{\Lambda}_{ac_k}}(\tilde{\gamma}_p|w)F_{\hat{\Lambda}_{c_kb}}(\tilde{\gamma}_p|w)$) under $\eta\rightarrow\infty$ as
	\begin{align}\label{ahsi}
	\tilde{\mathcal{P}}^{\textmd{sat}}_{\textmd{out}}(\mathcal{R}_{p})&=\mathbb{E}\{(F_{\hat{\Lambda}_{ac_k}}(\tilde{\gamma}_p|w)+F_{\hat{\Lambda}_{c_kb}}(\tilde{\gamma}_p|w))^K\}.
	\end{align}
	Then, as followed in \cite{pkss}, under $\eta\rightarrow\infty$, we simplify the cdfs $F_{\hat{\Lambda}_{ac_{k}}}(x|w)\simeq \frac{\alpha_{ac}x(w+1)}{\eta}$ and $F_{\hat{\Lambda}_{c_{k}b}}(x|w)\simeq\frac{1}{\Gamma(m_{cb}+1)}\left(\frac{m_{cb}x(w+1)}{\Omega_{cb}\eta}\right)^{m_{cb}}$, for small argument $x$. Upon inserting these cdfs into (\ref{ahsi}), we obtain 
	\begin{align}\label{asypp}
	\tilde{\mathcal{P}}^{\textmd{sat}}_{\textmd{out},\infty}(\mathcal{R}_{p})&=\left\{ \begin{array}{l}
	\left(\frac{\alpha_{ac} \tilde{\gamma}_p}{\eta}\right)^K \mathbb{E}\{(w+1)^K\},\textmd{ if } m_{cb} > 1, \\
	\left(\frac{\alpha_{ac} \tilde{\gamma}_p}{\eta}\!+\!\frac{\tilde{\gamma}_p}{\Omega_{cb}\eta}\right)^K \mathbb{E}\{(w+1)^K\}, \textmd{ if } m_{cb} = 1,
	\end{array}\right.
	\end{align}
	Finally, upon evaluating the expectation by following the similar steps as followed for $\mathcal{I}_1$ in Appendix \ref{appA}, we can achieve (\ref{asyp}).
	
	\section{Proof of Theorem \ref{tth2n}}\label{appC}
	We follow the initial steps similar to those in Appendix A, where the term $[\overline{F}_{\hat{\Lambda}_{c_kb}}(x|w)]^n$ for non-integer multinomial expansion is modified as 
	\begin{align}\label{mu3}
	[\overline{F}_{\hat{\Lambda}_{c_kb}}(x|w)]^n = &\sum_{v=0}^{n}\binom{n}{v}\frac{(-1)^v}{(\Gamma(m_{cb}))^v}\sum_{\overline{S}_p\in \overline{\mathcal{T}}}\frac{v!}{\prod_{p=0}^{\infty}\overline{s}_p!}\prod_{p=0}^{\infty}(\overline{\mathcal{B}}_p)^{\overline{s}_p}\left(x(w+1)\right)^{\overline{\Delta}_{cb}}.
	\end{align}
	
	Next, upon invoking (\ref{mu1}) and (\ref{mu3}) in (\ref{hrsi}), we have
	\begin{align}\label{intrB}
	\tilde{\mathcal{P}}^{\textmd{sat}}_{\textmd{out}}(\mathcal{R}_{p})&=\sum_{n=0}^{K}\binom{K}{n}(-1)^n\alpha^n_{ac}\sum_{S_{(\nu,m)}\in\mathcal{S}}\frac{n!}{\prod_{m=0}^{\varpi_{(\nu,ac)}}s_m!}\prod_{m=0}^{\varpi_{(\nu,ac)}}(\mathcal{A}_{(\nu,m)})^{s_m}\sum_{v=0}^{n}\binom{n}{v}\frac{(-1)^v}{(\Gamma(m_{cb}))^v}
	\\\nonumber
	&\times\sum_{\overline{S}_p\in \overline{\mathcal{T}}}\frac{v!}{\prod_{p=0}^{\infty}\overline{s}_p!}\prod_{p=0}^{\infty}(\overline{\mathcal{B}}_p)^{\overline{s}_p}\tilde{\gamma}^{\overline{\Delta}_{(\nu)}}_p\textmd{e}^{-{\Theta}_{(\nu,ac)} n \tilde{\gamma}_{p}}\underbrace{\mathbb{E}\{(w+1)^{\overline{\Delta}_{(\nu)}}\textmd{e}^{-{\Theta}_{(\nu,ac)} n \tilde{\gamma}_{p}w}\}}_{\mathcal{I}_2}
	\end{align}
	Upon taking the expectation $\mathcal{I}_2$ can be further represented as
	\begin{align}\label{68}
	\mathcal{I}_2&=\widetilde{\sum_{(\nu,s)}}\,\frac{\Xi_{(\nu,s)}(M_s)}{\eta^{\Lambda}_s}\left(\frac{m_t}{\Omega_t\eta_t}\right)^{m_t M_t}\frac{\Phi(m_t M_t,\Lambda)}{\Gamma(m_t M_t)}\int_{0}^{\infty}w^{\Lambda+m_tM_t-1}\left(w+1\right)^{\overline{\Delta}_{(\nu)}}\\\nonumber
	&\times\textmd{e}^{-\left({\Theta}_{(\nu,ac)}n\tilde{\gamma}_{p}+\frac{m_t}{\Omega_t \eta_t}\right)w}{}_1F_1(\Lambda;m_tM_t;-\tilde{\Theta}_{(\nu,s)}w)dw.
	\end{align}
	Unlike ${\Delta}_{(\nu)}$ in (\ref{63}), the term $\overline{\Delta}_{(\nu)}$ in (\ref{68}) may take non-integer values and thus, the binomial expansion for the term $\left(w+1\right)^{\overline{\Delta}_{(\nu)}}$ is not feasible. Hereby, we first resolve this problem by representing the function ${}_1F_1(\cdot;\cdot;\cdot)$ in series form \cite[eq. 9.210.1]{grad} and then, evaluating the resulting integral with the help of \cite[eq. 9.211.4]{grad} to obtain (\ref{asl23n}).

	\section{Proof of Corollary \ref{cor1n}}\label{appD}
	The proof follows the Appendix \ref{appB} where the case $m_{cb}=1$ is not applicable for noninteger values of $m_{cb}$. In particular, we can write the asymptotic OP as  
	\begin{align}\label{asypn11}
	\tilde{\mathcal{P}}^{\textmd{sat}}_{\textmd{out},\infty}(\mathcal{R}_{p})&=\left\{ \begin{array}{l}
	\left(\alpha_{ac} \tilde{\gamma}_p\right)^K\mathbb{E}\{(w+1)^K\},\textmd{ if } m_{cb} > 1, 
	\\\left(\frac{1}{\Gamma(m_{cb})}\right)^K\left(\frac{m_{cb}\tilde{\gamma}_p}{\Omega_{cb}}\right)^{m_{cb}K}\mathbb{E}\{(w+1)^{m_{cb}K}\},\textmd{ if } m_{cb} < 1.
	\end{array}\right.
	\end{align}
	Note that for $m_{cb}>1$, we can evaluate the resulting integral similar to $\mathcal{I}_1$ in Appendix \ref{appA}. However, for $m_{cb}<1$, the $m_{cb}K$ may be a noninteger quantity. So, we calculate the resulting integral in a similar manner as $\mathcal{I}_2$ in Appendix \ref{appC}.
	
	\section{Proof of Lemma \ref{lem1}}\label{appE}
	Based on (\ref{ous}), one can evaluate the conditional cdf
	\begin{align}\label{fst}
	{F}_{\hat{\Lambda}_{ac_{k^{\ast}}}}\!\!(x|w)
	\!&=\!K\textmd{Pr}\!\left[\hat{\Lambda}_{ac_k}\!\!<\!x, \Lambda_{ac_{k}b}\!>\!\max_{\underset{l\neq k}{l=1\dots K}} \Lambda_{ac_{l}b}|w\right].
	\end{align}
	Relying on the upper bound SINR $\Lambda^{\star}_{ac_{\jmath}b}$ for $\Lambda_{ac_{\jmath}b}$, $\jmath\in\{k,l\}$ as given by (\ref{hisn}), we can re-express (\ref{fst}) as 
	\begin{align}\label{fis}
	{F}_{\hat{\Lambda}_{ac_{k^{\ast}}}}(x|w)=K\textmd{Pr}\left[\hat{\Lambda}_{ac_k}<x, \digamma_k>\max_{\underset{l\neq k}{l=1\dots K}} \digamma_l|w\right],
	\end{align}
	where $\digamma_\jmath \triangleq  \min({\hat{\Lambda}_{ac_{\jmath}},\hat{\Lambda}_{c_{\jmath}b}})$. Now, we can evaluate (\ref{fis}) as 
	\begin{align}\label{fin}
	{F}_{\hat{\Lambda}_{ac_{k^{\ast}}}}(x|w)&=\varphi_1(x|w)+\varphi_2(x|w),
	\end{align}
	where
	\begin{align}\label{fin1}
	\varphi_1(x|w)&=K\int_{0}^{x}\int_{0}^{y}f_{\hat{\Lambda}_{ac_{k}}}(y|w)f_{\hat{\Lambda}_{c_{k}b}}(z|w)\left[F_{\digamma_k}(z|w)\right]^{K-1}dzdy
	\end{align}
	and 
	\begin{align}\label{fin2}
	\varphi_2(x|w)&=K\int_{0}^{x}f_{\hat{\Lambda}_{ac_{k}}}(y|w)\left(\int_{y}^{\infty}f_{\hat{\Lambda}_{c_{k}b}}(z|w)dz\right)\left[F_{\digamma_k}(y|w)\right]^{K-1}dy,
	\end{align}
	with 
	\begin{align}\label{mult}
	\left[F_{\digamma_k}(z|w)\right]^{K-1}&=\sum_{n=0}^{K-1}\binom{K-1}{n}(-1)^n\left[\overline{F}_{\hat{\Lambda}_{ac_k}}(z|w)\overline{F}_{\hat{\Lambda}_{c_kb}}(z|w)\right]^n
	\end{align}
	By first inserting the expressions (\ref{mu1}) and (\ref{mu2}) in (\ref{mult}) and then, utilizing the result in (\ref{fin1}) and (\ref{fin2}), one can obtain (\ref{fac}) after evaluating the integrals therein.
	
	\bibliographystyle{IEEEtran}
	\bibliography{IEEEabrv,bib}
	
\end{document}